\documentclass[11pt]{article}
\usepackage[IL2]{fontenc}
\usepackage[letterpaper,margin=1.00in]{geometry}
\usepackage{amsmath, amssymb, amsthm, amsfonts}
\usepackage{bbm}
\usepackage{amscd}
\usepackage{mathrsfs}

\usepackage{comment} 
\usepackage{ifthen}
\usepackage{tikz}
\usetikzlibrary{positioning,decorations.pathreplacing}
\usepackage{ bbold }
\usepackage{appendix}
\usepackage{graphicx}
\usepackage{color}
\usepackage{epstopdf}
\usepackage{wrapfig}
\usepackage{paralist}
\usepackage{wasysym}
\usepackage[textsize=tiny]{todonotes}

\usepackage{listings} 

\usepackage{pdflscape} 

\usepackage{framed}
\usepackage[framemethod=tikz]{mdframed}
\usepackage[bottom]{footmisc}
\usepackage{enumitem}
\setitemize{noitemsep,topsep=3pt,parsep=3pt,partopsep=3pt}
\usepackage[font=small]{caption}
\usepackage{xspace}

\usepackage{thmtools} 
\usepackage{thm-restate} 

\definecolor{darkgreen}{rgb}{0,0.5,0}
\definecolor{darkblue}{rgb}{0,0,0.6}
\usepackage{hyperref}
\hypersetup{
    unicode=false,          
    colorlinks=true,        
    linkcolor=darkblue,          
    citecolor=darkgreen,        
    filecolor=magenta,      
    urlcolor=cyan           
}

\newtheorem{theorem}{Theorem}[section]
\newtheorem{lemma}[theorem]{Lemma}
\newtheorem{meta-theorem}[theorem]{Meta-Theorem}
\newtheorem{claim}[theorem]{Claim}

\newtheorem{observation}[theorem]{Observation}
\newtheorem{definition}[theorem]{Definition}

\usepackage[capitalize, nameinlink,noabbrev]{cleveref}

\crefname{theorem}{Theorem}{Theorems}
\crefname{proposition}{Proposition}{Propositions}
\crefname{observation}{Observation}{Observations}
\crefname{lemma}{Lemma}{Lemmas}
\crefname{claim}{Claim}{Claims}
\crefname{problem}{Problem}{Problems}
\crefname{conjecture}{Conjecture}{Conjectures}
\crefname{question}{Question}{Questions}
\crefname{example}{Example}{Examples}
\crefname{fact}{Fact}{Facts}

\definecolor{darkgreen}{rgb}{0,0.5,0}

\usepackage{algcompatible}
\algnewcommand\algorithmicswitch{\textbf{switch}}
\algnewcommand\algorithmiccase{\textbf{case}}

\algdef{SE}[SWITCH]{Switch}{EndSwitch}[1]{\algorithmicswitch\ #1\ \algorithmicdo}{\algorithmicend\ \algorithmicswitch}%
\algdef{SE}[CASE]{Case}{EndCase}[1]{\algorithmiccase\ #1}{\algorithmicend\ \algorithmiccase}%
\algtext*{EndSwitch}%
\algtext*{EndCase}%

\newcommand{\eps}{\varepsilon}

\newcommand{\congest}{$\mathsf{CONGEST}\xspace$ }

\renewcommand{\P}{\textrm{P}}

\newcommand{\poly}{\operatorname{poly}}

\renewcommand{\phi}{\varphi}

\newcommand{\E}{\mathbb{E}}
\renewcommand{\Pr}{\P}

\renewcommand{\paragraph}[1]{\vspace{0.15cm}\noindent {\bf #1}:}

\newcommand{\FullOrShort}{full}
\ifthenelse{\equal{\FullOrShort}{full}}{
  
  \newcommand{\fullOnly}[1]{#1}
  \newcommand{\shortOnly}[1]{}

  }{

    \newcommand{\fullOnly}[1]{}
    \newcommand{\IncludePictures}[1]{}
   
  }

\newcommand{\Pot}

\renewcommand{\Pr}{\P}

\usepackage{algorithm}
\usepackage[noend]{algpseudocode}\usepackage{graphicx} 
\usepackage{amsmath}

\newcommand{\depth}{$\tilde{O}(\log^2 n)$}
\newcommand{\depthexact}{$O(\log^2 n \cdot \log \log n)$}
\newcommand{\round}{$\tilde{O}(\log n)$}
\newcommand{\roundexact}{$O(\log n \cdot \log \log n)$}
\newcommand{\roundexacttheta}{$\Theta(\log n \cdot \log \log n)$}

\title{Dynamic Graph Coloring: \\
Sequential, Parallel, and Distributed}
   \author{Mohsen Ghaffari \\ \small MIT \\ \small ghaffari@mit.edu
   \and
   Jaehyun Koo \\ \small MIT \\ \small koosaga@mit.edu}
\date{}

\begin{document}
\maketitle
\begin{abstract}
We present a simple randomized algorithm that can efficiently maintain a $(\Delta+1)$ coloring as the graph undergoes edge insertion and deletion updates, where $\Delta$ denotes an upper bound on the maximum degree. A key advantage is the algorithm's ability to process many updates simultaneously, which makes it naturally adaptable to parallel and distributed computations. Concretely, it gives a unified framework across the models, leading to the following results:

\begin{itemize}
    \item In the sequential setting, the algorithm processes each update in $O(1)$ expected time, \textit{worst-case}. This matches and strengthens the results of Henzinger and Peng [TALG 2022] and Bhattacharya et al. [TALG 2022], who achieved an $O(1)$ bound but \textit{amortized} over the sequence of updates (in expectation and with high probability, respectively), whose work improved the $O(\log \Delta)$ expected \textit{amortized} bound of Bhattacharya et al. [SODA'18]. 
    \item In the parallel setting, the algorithm processes each batch of updates using $O(1)$ work per update in the batch in expectation, and in $\poly(\log n)$ depth with high probability. This is, in a sense, an ideal parallelization of the above results. 
    \item In the distributed setting, the algorithm can maintain a coloring of the network graph as (potentially many) edges are added or deleted. The maintained coloring is always proper; it may become partial upon updates, i.e., some nodes may temporarily lose their colors, but quickly converges to a full, proper coloring. Concretely, each insertion and deletion causes at most $O(1)$ nodes to become uncolored, but this is resolved within \round~rounds with high probability (e.g., in the absence of further updates nearby--the precise guarantee is stronger, but technical). Importantly, the algorithm incurs only $O(1)$ expected message complexity and computation per update. 
\end{itemize}
\end{abstract}

\thispagestyle{empty}

\newpage
{
\tableofcontents

\thispagestyle{empty}
}
\newpage
\setcounter{page}{1}
\section{Introduction}
We present an efficient method for maintaining a $(\Delta+1)$-coloring in a dynamic graph that can process many updates simultaneously. Here $\Delta$ is (an upper bound on) the maximum degree. While a simple greedy algorithm solves the static $(\Delta+1)$-coloring problem in linear time sequentially, even this static case is challenging in parallel and distributed settings and has been the subject of much research. The difficulty grows further in the dynamic regime, where edges are continually inserted and deleted. We present a simple and unified framework for dynamic $(\Delta+1)$-coloring, which naturally adapts to each computational setting with near-optimal performance: 
\begin{itemize}
\item[(A)] In the sequential dynamic setting, it quantitatively matches the optimal $O(1)$ update time of the state of the art~\cite{henzinger2022constant,bhattacharya2022fully}, and qualitatively strengthens the update time bound from \textit{amortized} over the update sequence to expected per individual update (worst-case). This also leads to the first dynamic algorithm for (implicit) $(\Delta+1)$- coloring with $\poly(\log n)$ time per individual update, worst-case and with high probability. 
\item[(B)] In the parallel dynamic setting, it can process any batch of updates simultaneously using $O(1)$ expected computation per update in the batch, all in \depth{} depth. 
\item[(C)] In the distributed dynamic setting, it can process any batch of updates using $O(1)$ expected computation and communication per update in the batch, and the processing of each batch terminates in \round~rounds. The updates can arrive at arbitrary times, and do not need to wait for the previous batch's processing to finish.
\end{itemize}
The key ingredient is a set of simple randomized local repair rules that can handle updates efficiently, especially many \textit{concurrent} ones. Our overall scheme is arguably simpler than the (sequential) prior work, though the analysis is far more intricate, especially to achieve the expected bound per update. We next situate our results in their respective contexts. 

\subsection{Sequential Dynamic Coloring}
\textit{Dynamic graph algorithms} have been studied extensively over the past couple of decades. Besides addressing the needs arising in real-world applications, where graphs naturally undergo updates, they have proven essential even for static graph algorithms. In the latter, dynamic algorithms provide ``data structures" that efficiently update the solution to some subproblem, during the iterations of the broader static algorithm---see, e.g., \cite{chen2025maximum} for a celebrated recent example. Notable in this area is the rare breed of problems for which an \textit{$O(1)$ update time} has been achieved. Currently this includes constant-stretch (multiplicative) spanners~\cite{baswana2012fully}, maximal matching and vertex cover approximations~\cite{solomon2016fully}, and $(\Delta+1)$-coloring~\cite{bhattacharya2022fully,henzinger2022constant}. To the best of our knowledge, all these results are $O(1)$ \textit{amortized} update time, where the amortization/averaging is over the sequence of updates.

For dynamic coloring, the most relevant prior works are as follows: Bhattacharya, Chakrabarty, Henzinger, and Nanongkai~\cite{bhattacharya2018dynamic} gave a randomized dynamic algorithm that maintains a $(\Delta+1)$-coloring in $O(\log \Delta)$ amortized expected time per update. They also gave a deterministic algorithm that maintains a $(\Delta+o(\Delta))$-coloring in $\poly(\log \Delta)$ amortized update time. In subsequent work, Henzinger and Peng~\cite{henzinger2022constant} and independently Bhattacharya, Grandoni, Kulkarni, Liu, and Solomon~\cite{bhattacharya2022fully} gave randomized algorithms for $(\Delta+1)$-coloring with $O(1)$ \textit{amortized} time per update, in expectation and with high probability, respectively. 

As a side note, we comment that there has been much work also on dynamic algorithms for other variants of coloring, e.g., those depending on the graph's arboricity~ \cite{ghaffari2024dynamic,christiansen2023improved,solomon2020improved,henzinger2020explicit,barba2019dynamic} or for edge coloring~\cite{christiansen2023power, bhattacharya2024nibbling,duan2019dynamic,bhattacharya2018dynamic}. 

\paragraph{Our Sequential Result} Our simple dynamic algorithm matches the state-of-the-art constant-time per update in the sequential computation setting quantitatively, and strengthens the time bound guarantee qualitatively from amortized to expected per update (worst-case). Formally, it provides the following:

\begin{restatable}{theorem}{seqA}\label{thm:sequential-main}(\textbf{Sequential Worst-Case Dynamic Algorithm})
    There is a randomized dynamic algorithm that maintains an explicit $(\Delta + 1)$-coloring for a graph $G$ with $n$ vertices and maximum degree at most $\Delta$. The value $\Delta$ is fixed throughout the algorithm. The algorithm takes $O(n)$ time for initialization. The algorithm supports insertion and deletion of edges in $O(1)$ expected worst-case time against an oblivious adversary.
\end{restatable}
Our algorithm builds partially on the prior work~\cite{bhattacharya2018dynamic, bhattacharya2022fully, henzinger2022constant} but also has significant departures from them. We also point out a small flaw in the previous analysis, which appears easier to fix in the amortized setting but poses a more significant challenge in bounding the expected time per update. See \Cref{sec:MethodOverview} for more on these.

The qualitative strengthening from \textit{amortized} (expected or with high probability) $O(1)$ time, which means averaged over the sequence of updates, to \textit{worst-case} expected $O(1)$ time, which bounds the (expected) time per individual update, has other concrete implications: it enables us to apply a technique of Bernstein, Forster, and Henzinger~\cite{bernstein2019deamortization}, an implicit $(\Delta+1)$-coloring with polylogarithmic time per update, worst-case and with high probability. Concretely, we obtain a data structure  for $\Delta+1$ coloring, with worst-case $O(\log^2 n)$ update time with high probability, which has $O(1)$ time for each color query. This is the first result for $(\Delta+1)$-coloring with polylogarithmic \textit{worst-case update time}.\footnote{For implicit $O(\alpha)$-coloring, where $\alpha$ denotes the graph's arboricity, Ghaffari and Grunau~\cite{ghaffari2024dynamic} gave an algorithm with polylogarithmic \textit{worst-case} update time (and query), improving on an implicit $O(\min\{\alpha \log \alpha, \alpha \log\log\log n\})$-coloring of Christiansen, Nowicki, and Rotenberg~\cite{christiansen2023improved} that had polylogarithmic \textit{amortized} update time (and query).}

\begin{restatable}{corollary}{seqB}\label{cor:sequential-main}
    There is a randomized dynamic algorithm that maintains an implicit $(\Delta + 1)$-coloring for a graph $G$ with $n$ vertices and maximum degree at most $\Delta$. The value $\Delta$ is fixed throughout the algorithm. The algorithm takes $O(n)$ time for initialization. The algorithm supports insertion and deletion of edges in $O(\log^2 n)$ worst-case time w.h.p. against an oblivious adversary. The coloring is implicit: we do not update the list of colors explicitly, but each query for any vertex $v$ is answered in $O(1)$ time. 
\end{restatable}

\subsection{Parallel Batch-Dynamic Coloring}
The primary strength of our algorithm is its inherent adaptability to various computational models and its ability to handle multiple updates simultaneously. To make the statement precise, let us first recall the definitions for the parallel model.

\paragraph{Parallel Model---Static and Dynamic} We follow the standard work-depth terminology~\cite{jaja1992parallel, blelloch1996programming}. For an algorithm $\mathcal{A}$, its \textit{work} $W(\mathcal{A})$ is the total number of computations, and its \textit{depth} $D(\mathcal{A})$ is the longest chain of computations with sequential dependencies. By Brent's principle~\cite{brent1974parallel}, the time $T_p(\mathcal{A})$ for running the algorithm when we have $p$ processor can be tightly bounded with these parameters: $\max\{W(\mathcal{A})/p, D(\mathcal{A})\} \leq T_{p}(\mathcal{A}) \leq W(\mathcal{A})/p + D(\mathcal{A}).$ 

An emerging paradigm of \textit{batch-dynamic parallel algorithms} seeks to maximize and leverage parallelism in dynamic algorithms, by allowing large numbers of updates to be processed simultaneously. See, e.g., \cite{acar2019parallel,acar2020parallel,dhulipala2021parallel,tseng2022parallel,liu2022parallel,ghaffari2023nearly,anderson2024deterministic,ghaffari2024parallel, blelloch2025parallel,ghaffari2025parallelCoreness,ghaffari2025parallelSpanner}. This area lies at the intersection of \textit{dynamic sequential} and \textit{static parallel} algorithms, and mathematically, the results in it imply results in both of these settings. In this setting, graph updates arrive in (potentially large) batches, and the goal is to process each batch with low-depth computation and work proportional to the number of updates.

\paragraph{Static Parallel Algorithms for Coloring} A celebrated work of Luby~\cite[Section 6]{luby1985simple} gives a randomized static parallel algorithm that computes a $\Delta+1$ coloring. This was via a reduction to the maximal independent set problem, for which he gave an efficient parallel algorithm in the same paper. Luby originally stated the work bound simply as a polynomial, but a closer inspection reveals that (a minor adaptation of) the algorithm has $O(m+n)$ work in expectation, where $m$ denotes the number of edges. The algorithm's depth is $O(\log^2 n)$, with high probability. Historically, deterministic parallel algorithms had far worse bounds; but algorithms with $O((m+n)\poly(\log n))$ work and $\poly(\log n)$ depth are implicit in recent advances in deterministic distributed algorithms~\cite{ghaffari2022deterministic}.

\paragraph{Our Dynamic Parallel Result} Our dynamic coloring algorithm achieves a near-optimal parallelization of the result from the sequential dynamic setting. It can process each batch of updates with $O(1)$ expected work per update and in \depth{} depth with high probability. Notice that this, as a very special case, also implies a static parallel algorithm with $O(m+n)$ expected work and \depth{} depth, e.g., when all edges are inserted in one batch. 
\begin{restatable}{theorem}{pramA}\label{thm:pram-main}(\textbf{Parallel Batch-Dynamic Algorithm})
    There is a randomized batch-dynamic algorithm that maintains an explicit $(\Delta + 1)$-coloring for a graph $G$ with $n$ vertices and maximum degree at most $\Delta$. The value $\Delta$ is fixed throughout the algorithm. The algorithm takes $O(n)$ work and $O(1)$ depth for initialization. The algorithm supports batch insertion and deletion of edges in $O(1)$ expected worst-case work per edge in a batch, and \depthexact{} worst-case depth w.h.p for the entire batch. The algorithm works against an oblivious adversary.
\end{restatable}

\paragraph{Concurrent work} Hutton and Melrod~\cite{hutton2025parallel} recently gave a batch-dynamic parallel algorithm for $\Delta+1$ coloring. Our work was done independently and concurrently. Their algorithm also has polylogarithmic depth, but processes each batch using $O(\log \Delta)$ expected amortized work per update in the batch. Let us give a quantitative and qualitative comparison: their work bound is $O(\log \Delta)$ per update, while ours is $O(1)$. In addition, their work bound is \textit{amortized} (and in expectation), while ours is \textit{worst-case} (in expectation). Furthermore, their algorithm appears to be a parallel implementation of the approach of Bhattacharya et al.~\cite{bhattacharya2018dynamic}. Ours is a novel and simpler algorithm, which works in a uniform manner across the three models (sequential, parallel, and distributed) and achieves the optimal $O(1)$ worst-case expected work-per-update performance in all of these settings. 

\subsection{Distributed Batch-Dynamic Coloring}
\paragraph{Distributed Model---Static and Dynamic} We work with the standard message-passing model of distributed computation, with bounded-size messages (sometimes called \congest). The system consists of $n$ processors/computers, which are connected via a network modeled as an undirected graph $G=(V, E)$, with one node per processor. Per round, each processor/node can send one $O(\log n)$ bit message to each neighbor. At the end of the computation, each processor/node should have its own part of the solution, e.g., its color in the coloring problem. A primary measure of interest is the \textit{round complexity}, which is the number of rounds until all nodes terminate. Other important measures are \textit{communication complexity} (sometimes called \textit{message complexity}), which is the total number of messages sent during the algorithm, and \textit{computation complexity}, which is the total amount of computation over different processors. 

In the dynamic variant of the model, each round may bring an arbitrary batch of edge updates, which include edge insertions and deletions. Nodes can communicate through the newly inserted edges, but not through the deleted edges. These updates are specified distributedly, in the sense that each node receives the list of its newly inserted or deleted edges. The performance of the dynamic distributed algorithm is measured, in part, through its \textit{quiescence round complexity} and \textit{quiescence message complexity per update} (see, e.g., \cite{elkin2007near} and the citations therein), defined as follows: Suppose all graph changes stop occurring in some round $\alpha$, and that the distributed dynamic algorithm reaches the correct solution (in our case a full proper coloring) in round $\beta\geq \alpha$. The worst-case difference $\beta-\alpha$ is called the \textit{quiescence round complexity} of the algorithm. The worst-case number of messages sent during $\alpha-\beta$ is the \textit{quiescence message complexity} of the algorithm, and, noting that the number of updates can be different during different times, we particularly care about \textit{quiescence message complexity per update}. Frequently, when we are discussing a dynamic algorithm and that is clear from the context, we omit the word \textit{quiescence} and call these simply \textit{round complexity} and \textit{message complexity per update}.

\paragraph{Static Distributed Algorithms for Coloring} Distributed algorithms for $\Delta+1$ coloring have been the subject of much research. We mention some key results here. Luby's classic work~\cite{luby1985simple} gives a randomized algorithm with $O(\log n)$ round complexity, and it has $O((m+n))$ message complexity and computation, in expectation. Originally, this was via a reduction to the maximal independent set problem. Johansson~\cite{johansson1999simple} provided a direct and elegant phrasing of the algorithm. This $O(\log n)$ round complexity remained the state of the art for nearly three decades. Through a line of work~\cite{harris2016distributed,chang2018optimal,halldorsson2021efficient}, the randomized round complexity was reduced to $\poly(\log\log n)$. On the deterministic side, the state of the art had round complexity of $2^{O(\sqrt{\log n})}$ for a long time~\cite{awerbuch1989network,panconesi1992improved}. It was improved to $\poly(\log n)$ recently~\cite{rozhovn2020polylogarithmic}. See \cite{ghaffari2021improvedNC,ghaffari2023improvedNC} for improvements in the polylogarithmic exponent. With these static algorithms, any dynamic updates can be handled in a small number of rounds, by recomputing from scratch; however, the message complexity and the computation of this would be proportional to the entire graph and not just the updates.

\paragraph{Our Dynamic Distributed Result} Our dynamic algorithm adapts naturally to the distributed setting with batch-dynamic updates. It achieves $O(1)$ expected communication and computation complexity per update, and each batch's processing finishes within \round~rounds with high probability. Thus, the algorithm has $O(1)$ expected quiescence message complexity and computation complexity per update, and  \round~quiescence round complexity. Notice that, in particular, as a very special case for the static regime, this matches the performance of the classic algorithms of Luby~\cite{luby1985simple} and Johansson~\cite{johansson1999simple}. 
More formally, the algorithm achieves the following result.
\begin{restatable}{theorem}{congestA}\label{thm:congest-main}(\textbf{Distributed Batch-Dynamic Algorithm})
    There is a randomized batch-dynamic algorithm in a \textsf{CONGEST} model that maintains an explicit $(\Delta + 1)$-coloring for a graph $G$ with $n$ vertices and maximum degree at most $\Delta$. The value $\Delta$ is fixed throughout the algorithm. The algorithm supports batch insertion and deletion of edges in $O(1)$ expected message per edge in the batch, $O(1)$ expected computation (summed over all nodes) per edge in the batch, and \roundexact{} worst-case round w.h.p for the entire batch. The algorithm works against an oblivious adversary.
    
    More concretely, when batches of updates may arrive at arbitrary times, the
algorithm satisfies the following guarantees:
\begin{itemize}
    \item At the end of every round, the algorithm maintains a partial coloring
    of the graph, namely a proper coloring of an induced subgraph, while allowing
    some vertices to be temporarily uncolored.  We say that each uncolored vertex
    holds a \textbf{token}.  Each token has an owner, which is an edge inserted
    in some update batch.

    \item Each inserted edge creates at most two tokens, equivalently at most two
    temporarily uncolored vertices.  This bound holds deterministically, and the
    owner of each such token is the inserted edge.

    \item Edge deletions create no tokens.

    \item In each round, every token either remains at the same vertex, creates
    a new token at an adjacent vertex with the same owner, or disappears.

    \item For every inserted edge, all tokens owned by that edge disappear within
    \roundexact{} rounds after the edge is inserted, with high probability.
\end{itemize}
\end{restatable}


\paragraph{Other tangentially-relevant work on dynamic distributed algorithms} We are not aware of any prior work on distributed dynamic coloring, with polylogarithmic message per update and polylogarithmic round complexity.\footnote{Indeed, such a result would likely have implied a sequential dynamic coloring with polylogarithmic update time, thus providing an earlier result comparable to \cite{bhattacharya2018dynamic}. This is not an encompassing formal statement, just because the distributed model per se does not bound the local computations, and thus such an algorithm, despite having a small message complexity per update, could have had large computations.} Our result's strength especially lies in its \textit{$O(1)$ expected message complexity and computation per update}, which mirrors the $O(1)$ expected update time of the sequential dynamic algorithm. We next mention a few prior works tangentially related to distributed dynamic algorithms.

Censor-Hillel et al.~\cite{censor2016optimal} consider a dynamic maximal independent set in a much more restricted setting of dynamic distributed networks: their algorithm requires that updates arrive one by one, and each has enough time to be fully processed before the next update. Their algorithm recovers an MIS using $O(1)$ expected round complexity per update. However, the message complexity and computation can be arbitrarily large (and indeed even sequential dynamic MIS with sublinear and polylogarithmic update times were solved later~\cite{assadi2018fully,behnezhad2019fully,chechik2019fully}). Barenboim, Elkin, and Goldenberg~\cite{barenboim2021locally} gave a deterministic fully dynamic self-stabilizing $(\Delta+1)$-coloring algorithm with $O(\Delta+\log^{*}n)$ stabilization time after the last topology change or transient fault. Their protocol does not provide update-sensitive message or computation complexity and is incomparable to our result. Bamberger et al.\cite{bamberger2019local} consider $(\Delta+1)$ distributed coloring in a dynamic setting and show that the conflict potentially arising due to each edge insertion/deletion is resolved within $O(\log n)$ rounds. See their paper for the more nuanced definition of their dynamic guarantees. However, crucially, the algorithm uses up to $\Theta(n)$ messages and computation to handle a single insertion/deletion, and that makes it completely incomparable to our work (and indeed it did not imply anything about parallel or sequential dynamic algorithms for coloring). We note that if we cared simply about round complexity but not messages or computations, recomputing from scratch each time would take only $\poly(\log\log n)$ distributed rounds~\cite{chang2018optimal,halldorsson2021efficient}. 

Baswana et al. \cite{baswana2012fully} consider spanner constructions in distributed dynamic settings, and give an algorithm for constant-stretch multiplicative spanners with near-optimal size that has a constant amortized quiescence message complexity and computation per update, and $O(1)$ quiescence round complexity (the result applies to higher stretch, but the bounds scale exponentially). Barenboim~\cite{barenboim2016deterministic} considers $\Delta+1$ coloring in dynamic networks and gives a deterministic algorithm that, after the updates (and identifying and uncoloring nodes whose colors are improper now), recolors nodes in $\tilde{O}(\Delta^{3/4}) + O(\log^* n)$ rounds. Besides the large round complexity, the message complexity and computation per update can be very large (it is as if running the basic static $\tilde{O}(\Delta^{3/4}) + O(\log^* n)$-round algorithm on the graph induced by uncolored nodes, after gathering the colors of their neighbors).

\section{Method Overview}
\label{sec:MethodOverview}
In this section, we first present a brief overview of the methods in prior work~\cite{bhattacharya2018dynamic, bhattacharya2022fully, henzinger2022constant}. Then, we discuss our algorithm and its differences from the prior work. 

\subsection{Previous Methods}\paragraph{Bhattacharya et al. \cite{bhattacharya2018dynamic}, $O(\log \Delta)$ amortized expected update time} Bhattacharya et al. \cite{bhattacharya2018dynamic} use a dynamically maintained hierarchical partitioning of the vertices as the key structure in their dynamic algorithm. Vertices are placed in layers $1, 2, \ldots, \Theta(\log \Delta)$, and the following invariant is maintained, for a fixed constant $\beta>1$: each node $v$ in layer $\ell(v)$ has (1) at most $O(\beta^{\ell(v)})$ neighbors in layers $1$ to $\ell(v)$ and (2) at least $\Omega(\beta^{\ell(v)-5})$ neighbors in layers $1$ to $\ell(v)-1$. Said differently, it is crucial that the former number is upper-bounded by a constant factor of the latter. It is also helpful that the former numbers form a geometric series over different layers (this geometric series prevents losing an additional logarithmic factor in the time bound).

When a node $v$ in layer $\ell(v)$ needs to be recolored (due to a new edge insertion causing a color collision with a neighbor), node $v$ is not allowed to take any of the colors held by neighbors in equal or higher layers, i.e., layers $\ell\geq \ell(v)$. Each node $v$ always knows the colors used by neighbors in equal or higher layers; this is maintained with a data structure. Other colors are possible, subject to some further restrictions: Node $v$ picks a random color among the remaining colors for which either no neighbor in layers strictly below $\ell(v)$ has it (called a \textit{blank} color), or at most one such neighbor has it (called a \textit{unique} color). Using the invariant, one can see that at least $\Omega(\beta^{\ell(v)-5})$ such blank or unique colors exist. In the former case, node $v$ takes that blank color, and the process is done. In the latter case, assuming $w$ is the unique neighbor that has this unique color, node $v$ takes the color of node $w$, node $w$ loses its color, and we recursively color node $w$. It is crucial here that, thanks to the uniqueness, at most one such neighbor $w$ (who is also in a lower layer) has to recolor itself.

The recoloring step for node $v$ has cost proportional to $O(\beta^{\ell(v)})$. Since the chain proceeds to strictly lower layers, the total cost of the recolorings along the chain is a geometric series and bounded by $O(\beta^{\ell(v)})$. On the other hand, the color is chosen uniformly at random from a space of $\Omega(\beta^{\ell(v)-5})$ colors. Intuitively, this suggests that the probability of any particular future edge insertion creating a collision is $1/\Omega(\beta^{\ell(v)-5})$. Thus, the expected amortized cost would be bounded by $O(\beta^{\ell(v)})/\Omega(\beta^{\ell(v)-5}) = O(1)$. 

A warning about this probabilistic intuition: \textit{it is incorrect!} See \Cref{subsec:priorMethodCautions} for more on this. 

Of course, the crucial missing ingredient in this outline is a dynamic scheme for maintaining the hierarchical partitioning that satisfies the invariant for all nodes. Bhattacharya et al.~\cite{bhattacharya2018dynamic} have an elaborate mechanism for this, with an intricate potential function. The mechanism comes with an $O(\log \Delta)$ amortized time per update, thus dominating the overall update time.

\paragraph{Bhattacharya et al.~\cite{bhattacharya2022fully}, $O(1)$ amortized expected update time} The algorithm of Bhattacharya et al.~\cite{bhattacharya2022fully} merges the hierarchical partitioning above with the coloring mechanism, in a very careful manner. We do not discuss the details here as they do not relate to our algorithm, but it suffices to say that the dynamic movements in the hierarchical partitioning are done in a lazier way and are dependent on the coloring. The authors note, ``\textit{this [merge of the two mechanisms] makes the analysis ... significantly more challenging}''. Though we note that the ideal desired invariant and the base idea of the color sampling, namely choosing a random blank or unique color, remain the same. 

\paragraph{Henzinger and Peng~\cite{henzinger2022constant}, $O(1)$ amortized expected update time}
Henzinger and Peng do not use hierarchical partitioning. Their base structure is instead a random permutation of the nodes. Concretely, they give each node $v$ a uniformly random real number $r(v)\in [0,1]$, called the rank of $v$. When a node $v$ is to be recolored, it avoids the colors of higher-rank neighbors. It chooses among the remaining colors a color either not used by any neighbor (blank color) or at most one neighbor whose rank is at most the \textit{median} of the ranks of the neighbors that have rank below that of $v$ (unique color). In the latter case, assuming $w$ is the unique neighbor that holds that chosen random color, node $v$ gets colored, $w$ loses its color, and $w$ should be recolored. However, the scheme gets more involved at this point. On a high level, due to probabilistic dependencies in the rank analysis, the algorithm tracks which of the neighbors of $w$ were neighbors of $v$ (called old) and which were not (called new), and it bifurcates depending on which category has more than a constant fraction among the lower rank neighbors of $w$; it chooses random colors from different parts of the palette accordingly. More generally, in a chain of recolorings, the algorithm tracks which nodes were neighbors of previously visited nodes in the coloring chain. Notice that these are complications in the algorithm, not just the analysis. The analysis also gets complicated due to the possibility of \textit{bad} vertices that have a large share of previously visited neighbors, and the authors use a complex potential function to account for the cost incurred by the recoloring steps at such bad vertices. We note that, regardless, the ideal default would have been to take a random blank or unique color, as in Bhattacharya et al.~\cite{bhattacharya2018dynamic}, and these complications are just on top.

\subsection{Cautionary Notes about the Previous Methods}
\label{subsec:priorMethodCautions}
Let us add some words of caution about the methods in prior work. The goal is to explain, in part, why the previous algorithms seem inherently amortized over a sequence of updates, and why obtaining an expected bound for each individual update requires additional ideas.

\paragraph{Incorrectness of the intuitive color collision probability bound}
We mentioned above an intuitive probability bound: For each inserted edge $e={v,u}$, node $v$, which lies at layer $\ell(v)$, chose its color uniformly at random from a space of size $\Omega(\beta^{\ell(v)-5})$ at the time of its most recent recoloring. Thus, one might expect the probability that its color matches that of the other endpoint $u$ (which was colored before $v$) to be upper-bounded by $O(1/\beta^{\ell(v)-5})$. Taking this probability bound, or its analogue in the random-ranking setting, at face value might suggest that prior algorithms~\cite{bhattacharya2018dynamic, bhattacharya2022fully, henzinger2022constant} have a small expected running time for each individual update. However, this probability bound does not hold. The issue is subtle, and we also missed it in an early version of our work. For amortized update-time analysis, as achieved in~\cite{bhattacharya2018dynamic, bhattacharya2022fully, henzinger2022constant}, one can circumvent this issue.\footnote{To the best of our understanding,~\cite{bhattacharya2018dynamic} relied directly on this probability bound for each individual update, specifically in its proof of Lemma~3.5, and is therefore technically incorrect as written. The subsequent papers~\cite{bhattacharya2022fully, henzinger2022constant} do not rely on a per-update probability. Instead, they amortize over the update sequence by partitioning the execution into certain epochs and charging the complexities of different epochs against one another.} However, the resulting algorithms and analyses are fundamentally amortized and do not yield a bound on the expected processing time of each update. To clarify the point, we briefly explain the problem with the naive probability bound.


Even in the much simpler context of $((1+\eps) \Delta)$-coloring, the natural probability bound corresponding to the above does not hold. In $((1+\eps) \Delta)$-coloring, when a node $v$ chooses a color, at least $\eps\Delta$ colors are blank, and no neighbor has them-- this suggests that the probability that each edge insertion creates a collision is at most $O(1/(\eps \Delta))$. There is a counterexample to this, which we present in \Cref{app:counterexample}. We do that after briefly recalling the precise statements in the proofs of \cite{bhattacharya2018dynamic} and explaining the probabilistic issue involved. Our counterexample is a fixed update sequence (without knowing the randomness of the algorithm), in which the last edge insertion causes a color conflict with high probability. On a high level, the issue is this: in those analyses, at time $\tau$ of an edge insertion $e=\{v, u\}$, one looks back to the time $\tau_v$ of the last recoloring of $v$ and argues about the size of the space from which the color was chosen at that time. However, the value of $\tau_v$ itself is a random variable, and fixing it leaks information about the sampled color, namely that it was such that it did not cause a recoloring since time $\tau_v$. This skews the sampled color away from the colors of the endpoints of $v$-incident edges inserted after time $\tau_v$, and a priori, there is no control on the latter. 


\paragraph{Unique colors and their apparent necessity}
Note that all the previous methods, when recoloring a node, relied on the restriction to \textit{unique} colors besides the blank ones. That is, we want a color that at most one lower-priority neighbor (lower layer in the hierarchical partitioning~\cite{bhattacharya2018dynamic,bhattacharya2022fully} and lower rank in the random permutation~\cite{henzinger2022constant}) has. It would have been simpler, and arguably more natural, to not have such a restriction to \textit{unique} colors---instead, simply take a random color among those not taken by higher (or equal) priority neighbors (this is layers $\ell\leq \ell(v)$ in the hierarchical partitioning, and higher ranks in the permutation). But this may cause many lower-priority nodes to be forced into recoloring themselves. Moreover, since this is done iteratively, the number of those recolorings can grow rapidly. In the words of Bhattacharya et al.~\cite{bhattacharya2018dynamic}, this ``\textit{is a cascading effect which is hard to control.}" The unique color prevents a branching in the cascade, and it is the reason that all of the prior work~\cite{bhattacharya2018dynamic,bhattacharya2022fully,henzinger2022constant} used such a unique-color restriction. 

We note the issue is not merely simplicity: the uniqueness restriction makes the random color of node $v$ dependent on the color of lower-priority nodes. This makes the colorings have bi-directional dependencies on both lower-priority and higher-priority neighbors. That poses issues in the probabilistic analysis, especially when we want to bound the expected update time.

\subsection{Our Method}
Our algorithm, which achieves $O(1)$ expected time per individual update, certainly builds on the general coloring approach in the prior amortized work, but it has a number of important differences. We first explain the key differences. Afterward, we discuss the natural adaptation of our method to the distributed and parallel settings with batches of many simultaneous edge updates. 

\paragraph{Random hierarchical partitioning, from a geometric distribution} On the surface, we use the language of hierarchical partitioning. However, we do not compute the levels in any smart way, and we also do not update them dynamically. We simply assign to each node a random layer number drawn from a geometric series. Roughly speaking, $\Pr[l(v)=k]=2^{-k}$ for any integer $k\in \{1, 2, \ldots, \log \Delta+1\}$, with an appropriate rounding at the maximum.\footnote{Note that we have flipped the indexing of the layer numbers where higher layer suggests lower degree; this does not have any significant effect and is merely to match the notation in our technical writeup.} This assignment is done only once and never changed. The number of neighbors at the level $l(v)$ of node $v$ is in expectation $\Delta 2^{-l(v)}$ and about $\Theta(\Delta 2^{-l(v)})$ for levels strictly above $v$ (this statement assumes $v$ has degree about $\Delta$ but the problem gets considerably easier for $v$ if its degree is considerably below $\Delta$, as there will be more ``\textit{blank}" colors--here, for simplicity, we do not discuss the full generality). There can be deviations from these expectations, and the invariant desired above---specifically that the actual values of these are within a constant factor---might be broken for some nodes. However, this happens with a probability that is exponentially small in the number of neighbors in layer $\ell(v)$; one can formalize this using the memorylessness of the geometric distribution. This low probability enables us to absorb the cost incurred when we do not have the invariant, as the cost is only linear in the number of neighbors in the layer. 

In hindsight, the random level assignments in our algorithm mimic the elegant random ranks in the algorithm of Henzinger and Peng~\cite{henzinger2022constant}, placing it within the hierarchical partitions of Bhattacharya et al.~\cite{bhattacharya2022fully, bhattacharya2018dynamic}. Our algorithm does not need to track previously visited neighbors along the coloring chain and bifurcate according to their prevalence, as done by Henzinger and Peng. This makes the algorithm arguably much simpler and allows us to handle multiple updates simultaneously. 

\paragraph{Removing the uniqueness restriction} Another important change we have is that we let go of the restriction to unique colors. When a node $v$ at level $l(v)$ is to be recolored, it samples a color uniformly at random from those not taken by neighbors in layers $\ell\leq l(v)$. As noted by prior work, the challenge is that this can cause a problem for several neighbors in layers $\ell'>l(v)$. We simply make them all recolor and show, only in the analysis, that the total expected work remains bounded. In short, the base observation is this: one can show that, in expectation, at most one neighbor is forced to recolor. However, working only with an expected bound and dealing with the possibility of the recolorings branching makes the analysis more intricate. We will need stronger concentration analyses for the costs of various nodes reached in this stochastic branching process of recoloring nodes. On the positive side, this liberty (of choosing any color not taken by neighbors in $\ell\leq l(v)$) enables us to show a simple-yet-powerful independence between the colors: the color of a vertex at level $l(v)$ is determined by random choices made at levels at most $l(v)$, and is independent of random choices made at strictly higher levels. This independence significantly simplifies the task of analyzing the expected cost per update. 

\paragraph{Proactive probabilistic recolorings} Furthermore, besides the recolorings triggered by monochromatic edge insertions, we also have two types of additional probabilistic recolorings: (1)
Upon each insertion incident on a node $v$, even if it does not cause a color conflict, we recolor node $v$ with probability $\Theta(1/(\Delta 2^{-\ell(v)}))$. (2) Upon each node $u$ recoloring itself, for each neighbor $v$ of $u$ with $\ell(v)>\ell(u)$, even if the recoloring did not cause a conflict, we recolor node $v$ with probability $\Theta(1/(\Delta 2^{-\ell(v)}))$. These are crucial to our analysis, and especially to proving a bound on the probability of each edge insertion causing a color conflict. In an informal sense, they enable us to argue that it is unlikely that node $v$ has had many edge insertions since its last time of recoloring or many neighbors $u$ with $\ell(u)<\ell(v)$ that have recolored themselves. As such, we are able to provide bounds on the number of potentially ``relevant" recolorings that node $v$ has had and use that to bound the probability of color collision for each edge insertion. Regrettably, this part of the analysis is quite involved, but that appears necessary to have a correct formalization of the intuitive probability bound, and to prove the $O(1)$ expected time per individual update.
Moreover, because of these proactive probabilistic recolorings, the branching process of which neighbors get uncolored grows even faster and has an expected out-degree exceeding one, though with the fortunate property that most are in higher-numbered layers, which have smaller expected relevant degree, and thus we can bound the overall work caused by the whole branching process.

\paragraph{A brief summary of the algorithmic differences} To summarize, the key differences between our sequential algorithm with those of the prior work can be listed as follows: (1) we use simple random layers from a geometric distribution instead of the dynamically maintained hierarchical partitioning, (2) we do not have the restriction to unique colors and instead analyze the more complicated stochastic branching process of recolorings, (3) we have additional probabilistic recolors besides the basic collision-triggered ones, and this enables us to analyze the probability of color collision even through the complex process of colors changing with dependencies on each other.  


\paragraph{Parallel and distributed adaptations} Our dynamic algorithm is made of a simple randomized layering and basic local recoloring rules. It naturally lends itself to handling multiple updates simultaneously and thus enables us to obtain near-optimal batch-dynamic algorithms for coloring in the parallel and distributed settings mentioned earlier. The details in these settings need further explanation, but on a high level, we can have different uncolored nodes try their colorings simultaneously. Each node to be recolored erases its previous color, and they all simultaneously attempt new colors at random. There are tie-breaking mechanisms between neighbors, depending on their layers. We show that each node succeeds with a constant probability to get colored, per iteration. However, when this happens for a node $v$, it can make some of its neighbors in layers $\ell>l(v)$ uncolored (by color collisions or proactive recolorings), and they have to do their own randomized colorings. Regardless, we are able to analyze these (simultaneous) stochastic branching processes of the uncolored nodes. We show that the expected work bound per insertion/deletion remains $O(1)$, and furthermore, for each batch of insertions or deletions (with an arbitrary number of insertions or deletions), within \roundexact{} iterations of color samplings, w.h.p., all nodes are colored; equivalently, no node remains uncolored.

\section{Preliminaries}
\paragraph{Number of Updates} Following standards in the area, we assume that the number of updates in the lifetime of the dynamic algorithm is at most a (large) polynomial in the number of vertices $n$. This is because we use certain data structures (e.g., hash functions) that have a $1/\poly(n)$ failure probability, and we can union-bound this failure over all updates. In all cases of our algorithm, the failures are detectable. We emphasize that in our sequential and parallel algorithms, we can easily remove the restriction on the number of updates: if a failure happens, we restart the entire dynamic algorithm (and all the randomness used). Since the initialization has a cost of at most $\poly(n)$ and the failure occurs with a probability of at most $1/\poly(n)$ per update, this changes the expected cost per update by only an additive $O(1)$. 


\paragraph{Random Primitives} In the pseudo-code for the algorithms, $\textsc{WithProbability}(p)$ takes a real number $p \in [0, 1]$ and returns $\textsf{TRUE}$ with probability $p$ and $\textsf{FALSE}$ with probability $1-p$.

\paragraph{Hash Tables} We use the parallel randomized hash table introduced in Gil et al. ~\cite{gil1991towards} for maintaining a dictionary. The parallel randomized hash table of \cite{gil1991towards} supports batch insertion and batch deletion of elements, each of them being a pair of key and value. The hash table supports batch find queries, which means that for each query, it finds the element that matches the query key, or states that such an element does not exist. The algorithm uses $O(1)$ work per update or query, and $O(\log n)$ depth for each batch operation. The work and depth bounds hold with high probability.

\paragraph{Concentrations} We use the Chernoff bound, stated below:
\begin{theorem}[Chernoff's bound]\label{thm:chernoffva}
For independent Bernoulli variables $X_1, \ldots, X_n$, let $X = \sum_{i = 1}^n X_i$ and $\mu = E[X]$. Then, for any $\eps \in (0, 1)$ we have $\Pr[X \geq (1+\eps) \mu] \leq \exp(\frac{-\eps^2 \mu}{3})$, and for any $\eps \in (0, 1)$ we have $\Pr[X \leq (1 - \eps) \mu] \leq \exp(\frac{-\eps^2 \mu}{2})$. In addition, for any $\eps\geq 1$, we have $\Pr[X\geq (1+\eps)\mu] \leq \exp(\frac{-\eps \mu}{3})$.
\end{theorem}

\paragraph{Correlation Inequalities} The following inequality is useful for the analysis. Let $\Omega = \Omega_1 \times \Omega_2 \times \cdots \times \Omega_n$ be a sample space, and suppose that each $\Omega_i$ is totally ordered. For each $x = (x_1, x_2, \ldots, x_n) \in \Omega$ and $y = (y_1, y_2, \ldots, y_n) \in \Omega$, we say that $x \leq y $ if and only if $x_i \leq y_i$ for all $1 \le i \le n$. We call an event $A \subseteq \Omega$ \textit{increasing} if $x \in A, x \le y$ implies $y \in A$ (for all $y\geq x$). Similarly, we call an event $A \subseteq \Omega$ \textit{decreasing} if $x \notin A, x \le y$ implies $y \notin A$ (for all $y\geq x$). Then, Harris~\cite{harris1960lower} proved the following inequality (this is a special case of the FKG inequality--see, e.g., \cite[Chapter 6]{alon2016probabilistic}):

\begin{theorem}[Harris' Inequality]\label{thm:harris}
If \(A\) and \(B\) are both increasing events of independent random variables, then
\[
  \Pr[A \cap B] \ge \Pr[A]\,\Pr[B].
\]

The same results hold even if \(A\) and \(B\) are both decreasing events of independent random variables.
\end{theorem}

\section{Definitions and Data Structures}
We assume that the graph contains no loops and duplicate edges throughout the algorithm, and any updates that violate these requirements are considered invalid. Our algorithm is guided by a randomly assigned hierarchy among the vertices. We assign each vertex $v \in V$ the following attributes:

\begin{itemize}
    \item A \textit{level} $l(v)$ is sampled independently from a geometric distribution. Concretely, $\Pr[l(v)=k]=2^{-k}$ for all $k\in [1, \lceil\log \Delta\rceil]$, and $\Pr[l(v)=k]=2^{-k+1}$ for $k=\lceil\log \Delta\rceil+1$.
    \item A \textit{color} $c(v)$ is obtained by independently sampling a random number from the set $\{1, \dots, \Delta+1\}$.  Our objective is to maintain a proper coloring, where for any edge $(u,v)$, we have $c(u) \neq c(v)$. Clearly, the initial coloring is proper, as the graph has no edges.
    \item A \textit{timestamp} $t(v)$ is the last time at which the vertex was assigned a color. The time corresponds to the sequential index of each unit of function calls. $t(v)$ is always distinct -- if we assign colors to multiple vertices simultaneously (either in the initialization stage, or in a parallel / distributed case), we assign $t(v)$ to be an arbitrary distinct value that is consistent with the previous assignments (in other words, any newly assigned $t(v)$ is greater than any previously assigned values of $t(v)$).
\end{itemize}

Following the definition of a level, we define degrees that count only neighbors at most/at least the vertex’s level:

\begin{definition}
    For a vertex $v$, we use $NB_\leq(v)$ to denote the set of neighbors of $v$ that have level at most $l(v)$. We define $NB_<(v), NB_>(v), NB_\geq(v)$ similarly. We use $d_\leq(v) = |NB_\leq(v)|$ to denote the number of neighbors of $v$ that have level at most $l(v)$. We define $d_<(v), d_>(v), d_\geq(v)$ similarly. 
\end{definition}

We additionally define the following:
\begin{definition}
    A color is \textit{available} for $v$ if it does not occur on any neighbors of level at most $l(v)$. We use use $\mathcal A_v$ to denote the set of all available colors for $v$.
\end{definition}

Each vertex $v$ keeps the following data structures:
\begin{itemize}
    \item \(\textsc{LowerEqualUsed}(v)\): A randomized data structure that
    stores the multiset of colors used by neighbors of \(v\) whose level is at
    most \(l(v)\).  It supports insertion (\(\textsc{Insert}\)), deletion
    (\(\textsc{Delete}\)), and \(\textsc{SampleNotIn}\).  The operation
    \(\textsc{SampleNotIn}\) samples uniformly at random from the set of all
    colors in \([1,\Delta+1]\) whose occurrence count in the multiset is zero.
    Each operation takes \(O(1)\) expected time.  In the parallel algorithm, the
    data structure supports batch operations with \(O(1)\) worst-case work per
    element and \(O(\log n)\) worst-case depth for the entire batch.
    \item $NB_{\geq}(v)$: A hash set of neighbors that have their level greater than or equal to that of $v$.
    \item $NB_{>}(v)$: A hash set of neighbors that have their level greater than that of $v$.
\end{itemize}

Note that the data structure $\textsc{LowerEqualUsed}$ stores exactly the complement of $\mathcal A_v$, and hence will support sampling from $\mathcal A_v$ in $O(1)$ expected time. 

We finish this section by presenting the implementation of $\textsc{LowerEqualUsed}$. Before doing so, we show a data structure that supports uniform sampling from the set. Our parallel batch-dynamic implementation is work-efficient and supports $O(1)$ worst-case operation, which is also sufficient for our sequential algorithm.

\begin{lemma}\label{lem:jhnah}
Suppose that we are dynamically maintaining a multiset $S\subseteq [1, \Delta + 1]$. There is a randomized batch-dynamic data structure that supports batch insertion and deletion of elements in $S$, as well as sampling an element chosen uniformly at random from the set of distinct elements currently in $S$. Every sampling call returns a newly sampled element, independent of previous samples. Each distinct element is sampled with uniform probabilities, and the number of its occurrences is ignored. The algorithm supports each operation in $O(1)$ worst-case work per element in a batch and $O(\log n)$ worst-case depth for a whole batch w.h.p. The algorithm works against an oblivious adversary.
\end{lemma}
\begin{proof}
    Assume that $S$ is a set of distinct elements. This can be accomplished by using an additional batch-dynamic hash table that maps each element to its occurrence count and ignores updates that do not change the occurrence count between zero and non-zero.

    We use two hash tables $H$ and $H_{rev}$, where $H$ maps each index of $0, 1, \ldots |S| - 1$ to a distinct element in $S$, and $H_{rev}$ stores the inverse mapping.

    \begin{itemize}
        \item To sample a random number, we sample a random index $i \in [0, |S| - 1]$ and return the element of index $i$ in $H$.
        \item To insert a set of numbers $B$ in a set $S$ with $n$ elements, we first ensure $B \cap S = \emptyset$ using the batch membership query. Then, we label each element of $B$ with a sequential index from $n, n + 1, \ldots$, which takes $O(|B|)$ work and $O(\log |B|) = O(\log n)$ depth. Finally, using the batch insertion, we put each element in $B$ with the corresponding index in $H$, and the reverse mapping in $H_{rev}$.
        \item To delete a set of numbers $B$ in a set $S$ with $n$ elements, we first ensure $B \subseteq S$ using the batch membership query. Let $k = |B|$. Using the batch membership query, we split $B$ into two parts $ B_l$ and $ B_r$, where $B_l$ contains all elements in $B$ whose index in $H$ is less than $ n-k$, and $B_r = B \setminus B_l$. We first remove all elements in $B$ from $H$, and for each index $n - k, n-k+1, \ldots, n-1$ that is vacated by the removal of $B$ in $H$, we map them arbitrarily with the elements in $B_l$. Note that $|B_l| = k - |B_r|$, so such a mapping is possible. Finally, for all elements in $H$ with index at least $n - k$, move to the mapping position, which has an index less than $n-k-1$ and is vacated by the removal of $B$. Finally, apply all the changes to the reverse mapping in $H$. We can implement this procedure with sequential indexing of a list, which, as in the insertion case, takes $O(|B|)$ work and $O(\log |B|) = O(\log n)$ depth.
    \end{itemize}

    Note that $H$ only requires the functionality of a dynamic array. The result follows from the batch-dynamic hash tables of \cite{gil1991towards}. 
\end{proof}

We now present our implementation of $\textsc{LowerEqualUsed}$. The straightforward way to implement $\textsc{LowerEqualUsed}$ is to initialize the data structure of \cref{lem:jhnah} with $[1, \Delta + 1]$ and invert the operation (delete the element in the insertion update, and vice versa). However, this will require a $O(n \Delta)$ initialization time, which can be large depending on the application. In \cref{lem:inversesample}, we take an extra effort to avoid this overhead. 

\begin{lemma}\label{lem:inversesample}
Suppose that we are dynamically maintaining a set $S\subseteq [1, \Delta + 1]$. There is a randomized batch-dynamic data structure that supports batch insertion and deletion to the set, as well as sampling a random number in $[1, \Delta + 1] \setminus S$. Every sampling call gives a newly sampled element, independent of the previous sample. The algorithm supports insertion and deletion operations in $O(1)$ worst-case work per element in a batch and $O(\log n)$ worst-case depth for a whole batch w.h.p. The algorithm supports the sampling operation in $O(1)$ expected worst-case work and $O(\log n)$ worst-case depth w.h.p. The algorithm works against an oblivious adversary.
\end{lemma}
\begin{proof}
    Assume that $S$ is a set of distinct elements. This can be accomplished by using an additional batch-dynamic hash table that maps each element to its occurrence count and ignores updates that do not change the occurrence count between zero and non-zero.

    We maintain the $S$ as a hash set. We additionally maintain the complement of $S$, $S^C$, using the data structure of \cref{lem:jhnah}, but only if the size of $S$ is sufficiently large (and hence the size of $S^C$ is sufficiently small). We also assume $\Delta + 1\geq 100$ --- otherwise, we resort to a sequential list.

    For maintaining the set $S^C$, our goal is to slowly populate it so that $S^C$ is empty if $|S| \leq \frac{1}{2}(\Delta + 1) - 2$, and $S^C$ is correctly populated if $|S| \geq \frac{3}{4}(\Delta + 1)$. Let $f(n) = \min(\Delta + 1, 4(n - \lceil \frac{1}{2}(\Delta + 1) \rceil))$. If the size of $S$ is $n$, we want $S^C$ to be equal to the set $[1, f(n)] \setminus S$. For this, when inserting into a set of size $k$ into a set of size $n$, or deleting a set of size $k$ from a set of size $n+k$, we check the updated element, and all integers in the range $(f(n), f(n + k)]$, and update their status correctly to the set $S^C$. For any choice of $n$, this amounts to a batch update of size $O(k)$ in the set $S^C$, and by \cref{lem:jhnah}, we can do this in $O(1)$ worst-case work per element and $O(\log n)$ total depth w.h.p.
    
    To sample a random number, if $|S| \leq \frac{3}{4}(\Delta + 1) + 5$, we sample a random number from $[1, \Delta + 1]$ until we find the one that does not belong to $S$. This procedure will end in expected $O(1)$ worst-case work. This procedure will end in $O(\log n)$ depth w.h.p. Otherwise, we have $f(|S|) = \Delta + 1$, so we use the data structure $S^C$ to sample a random number in $S^C$.
\end{proof}

\section{Sequential Dynamic Algorithms}\label{sec:sequential}
In this section, we prove the following theorem:
\seqA*

As mentioned in the introduction, the worst-case expected bound of the above result enables us to achieve an implicit coloring with polylogarithmic worst-case update time, with high probability, as formally stated in \Cref{cor:sequential-main}. Assuming \cref{thm:sequential-main}, we prove \cref{cor:sequential-main} using the reduction of Bernstein et al.~\cite{bernstein2019deamortization}. 
\seqB*
\begin{proof}[Proof of \Cref{cor:sequential-main}]
We maintain $\Theta(\log n)$ copies of the data structure from \cref{thm:sequential-main}, which we denote $D_1, D_2, \ldots, D_{\Theta(\log n)}$. Each copy will use independent randomizations.
For each update, we issue the same update to each copy of the data structure $D_i$. However, each $D_i$ will not perform the update directly, but instead put the update into the back of the \textit{computation queue}. Then, each copy will perform $\Theta(\log n)$ computations starting from the front of the computation queue, stopping if the number of computations has exceeded $\Theta(\log n)$ or the queue is empty. In \cite[Theorem 1.1]{bernstein2019deamortization}, Bernstein et al. prove the following: Suppose that each $D_i$ takes $O(1)$ worst-case expected time for each update. Then, after each update, there is at least one data structure copy with an empty computation queue, with high probability. We use this copy to answer color queries. Since we assume that the number of updates is at most $\poly(n)$, the correctness holds w.h.p. by the union bound.
\end{proof}

\subsection{Algorithm}
After every update, we maintain the data structures $\textsc{LowerEqualUsed}$, $NB_{\geq}$, and $NB_{>}$ in $O(1)$ expected time. For an edge deletion, this is the only work performed.

For each vertex $v$, define

\[
\theta(v)=\left\lceil \frac{\Delta\cdot 2^{-l(v)}}{100}\right\rceil.
\]

Observe that $\theta(v)$ is within a constant factor of the expected number of neighbors whose level is strictly larger than $l(v)$.

Consider an edge insertion $(u,v)$. If the insertion creates a color collision, i.e., $c(u)=c(v)$, then we recolor one of the endpoints. Specifically, we recolor the endpoint that is larger in the lexicographic order induced by $(l(\cdot),t(\cdot))$, where $l(\cdot)$ denotes the level and $t(\cdot)$ the timestamp. Afterward, each endpoint $w\in\{u,v\}$ is \emph{refreshed}: independently, with probability $\frac{10^{-4}}{\theta(w)}$, we invoke $\textsc{Recolor}(w)$.

The recoloring procedure for a vertex $v$ is as follows. We sample a uniformly random color from $\mathcal A_v$ and assign it to $v$. By construction, the new color does not conflict with any neighbor whose level is at most $l(v)$. However, conflicts may remain with neighbors at higher levels. Whenever such a conflict occurs, we recursively recolor the higher-level neighbor.

In addition, every higher-level neighbor of $v$ is refreshed after the recoloring. Consequently, a recursive recoloring may be triggered for the higher-level neighbor even in the absence of a collision.

We emphasize that the recolorings are not required to follow any global order (e.g., level or timestamp order), and the same vertex may be recolored multiple times during a single update.

We present the pseudocodes below.

\begin{algorithm}[H]
\caption{Procedure $\textsc{InsertEdge}(u, v)$}\label{alg:insertedge}
\begin{algorithmic}[1]
\small
\State Update $\textsc{LowerEqualUsed}, NB_\geq, NB_>$.
\If{$(l(u),t(u))>_{\mathrm{lex}}(l(v),t(v))$} \Comment{Lexicographic comparison of pairs}
\State $\textsc{Swap}(u, v)$
\EndIf
\If{$c(u) = c(v)$}
\State $\textsc{Recolor}(v)$
\EndIf
\State $\textsc{Refresh}(u)$
\State $\textsc{Refresh}(v)$
\end{algorithmic}
\end{algorithm}
\begin{algorithm}[H]
\caption{Procedure $\textsc{DeleteEdge}(u, v)$}\label{alg:deleteedge}
\begin{algorithmic}[1]
\small
\State Update $\textsc{LowerEqualUsed}, NB_\geq, NB_>$.\end{algorithmic}
\end{algorithm}
\begin{algorithm}[]
\caption{Procedure $\textsc{Refresh}(v)$}\label{alg:refresh}
\begin{algorithmic}[1]
\small
\If {$\textsc{WithProbability}(\frac{10^{-4}}{\theta(v)})$}
\State $\textsc{Recolor}(v)$
\EndIf
\end{algorithmic}
\end{algorithm}
\begin{algorithm}[H]
\caption{Procedure $\textsc{Recolor}(v)$}\label{alg:recolor}
\begin{algorithmic}[1]
\small
\State Update the timestamp $t(v)$ to the current time.
\For{$w \in NB_{\geq}(v)$}
\State $\textsc{LowerEqualUsed}(w).\textsc{Delete}(c(v))$
\EndFor
\State $c(v) = \textsc{LowerEqualUsed}(v).\textsc{SampleNotIn}()$
\For{$w \in NB_{\geq}(v)$}
\State $\textsc{LowerEqualUsed}(w).\textsc{Insert}(c(v))$
\EndFor
\For{$w \in NB_{>}(v)$}
\If {$c(v) = c(w)$}
\State $\textsc{Recolor}(w)$
\EndIf
\State $\textsc{Refresh}(w)$
\EndFor
\end{algorithmic}
\end{algorithm}


\subsection{Collision Probability}\label{sec:step1}
Throughout this section, we treat the level assignment as fixed and arbitrary. In particular, the arguments in this section do not use any randomness from the level assignment.

A key property of the algorithm is that the behavior of a vertex $v$ is not affected by random choices made at vertices of a higher level.

\begin{observation}\label{obs:hierarchy}
Fix the graph, the update sequence, and the level assignment. At any given time, the color of a vertex $v$ is a deterministic function of these fixed objects together with the random choices made at vertices of level at most $l(v)$. More precisely, if we know all random choices made by calls to $\textsc{Refresh}(w)$ and $\textsc{LowerEqualUsed}(w).\textsc{SampleNotIn}()$ over all vertices $w$ with $l(w)\le l(v)$, then the color of $v$ is determined. In particular, the color of $v$ is independent of all random choices made at vertices whose level is strictly larger than $l(v)$.
\end{observation}

We prove two lemmas. \cref{lem:jump} handles the collision probability for an update that connects two vertices of different levels, while \cref{lem:jump2} handles the collision probability for an update that connects two vertices of the same level. Both proofs require some care with the algorithm's low-level behavior. After proving these lemmas, we show that they imply the collision bound required by the actual algorithm.

\begin{definition}\label{def:goodlevel}
    Fix the graph, the update sequence, a vertex $v$, and its level $l(v)$. We call a level assignment \textbf{good} for $v$ if it satisfies $(\Delta+1-d_{\leq}(v))\ge \theta(v)$ throughout the last $2\theta(v)^2$ edge insertions incident to $v$. Otherwise, we call the level assignment \textbf{bad} for $v$. This property is determined solely by the input sequence and the level assignment. In particular, it is independent of the random choices made by the recoloring procedure.
\end{definition}

\begin{lemma}\label{lem:jump}
Fix the following objects in order, where each object may depend on the preceding ones:
\begin{itemize}
\item the graph, initially without edges;
\item the update sequence;
\item the level assignment;
\item a vertex $v$ such that the level assignment is good regarding $v$;
\item all random choices made by calls to $\textsc{Recolor}$ and $\textsc{Refresh}$ at vertices whose level is less than $l(v)$f;
\item a color $q^* \in[\Delta+1]$.
\end{itemize}
Then, over the remaining random choices made by calls to $\textsc{Recolor}$ and $\textsc{Refresh}$ at vertices whose level is at least $l(v)$, the probability that $c(v)=q^*$ at the end of the update sequence is at most $O(1/\theta(v))$.
\end{lemma}

\begin{lemma}\label{lem:jump2}
Fix the following objects in order, where each object may depend on the preceding ones:
\begin{itemize}
\item the graph, initially without edges;
\item the update sequence;
\item the level assignment;
\item a vertex $u$;
\item a vertex $v$ such that the level assignment is good regarding $v$;
\item all random choices made by calls to $\textsc{Recolor}$ and $\textsc{Refresh}$ at vertices whose level is less than $l(v)$.
\end{itemize}
Then, over the remaining random choices made by calls to $\textsc{Recolor}$ and $\textsc{Refresh}$ at vertices whose level is at least $l(v)$, the probability that $c(u)=c(v)$ and $t(u)<t(v)$ at the end of the update sequence is at most $O(1/\theta(v))$.
\end{lemma}
We first show some observations that will be used in the proofs of both lemmas. Define the following two types of events that are \emph{relevant} to $v$:
\begin{itemize}
\item \textbf{Type A:} an edge incident to $v$ is inserted;
\item \textbf{Type B:} a neighbor of $v$ whose level is less than $l(v)$ resamples its color in line 4 of \cref{alg:recolor}.
\end{itemize}

For both lemmas, the fixed context determines the full sequence of events relevant to $v$. Type A events are determined directly by the update sequence. Type B events are determined by \cref{obs:hierarchy}, since they involve only vertices of level less than $l(v)$. Let $R_1, R_2, \ldots, R_k$ denote the resulting sequence of relevant events in the order of occurrence.

Every recoloring of $v$ in line 4 of \cref{alg:recolor} arises from exactly one of the following cases:
\begin{itemize}
\item An edge incident to $v$ is inserted, and $\textsc{Recolor}(v)$ is invoked directly. This event can occur only during a Type A event.

\item An edge incident to $v$ is inserted, causing $\textsc{Refresh}(v)$ to be invoked, and this call to $\textsc{Refresh}(v)$ invokes $\textsc{Recolor}(v)$. This event occurs exactly when a Type A event occurs and the call to $\textsc{WithProbability}$ in line 1 of \cref{alg:refresh} returns $\textsf{TRUE}$.

\item An edge not incident to $v$ is inserted, and $\textsc{Recolor}(v)$ is invoked as part of a recursive recoloring chain. This event can occur only during a Type B event.

\item An edge not incident to $v$ is inserted, line 11 of \cref{alg:recolor} invokes $\textsc{Refresh}(v)$, and this call to $\textsc{Refresh}(v)$ invokes $\textsc{Recolor}(v)$. This event occurs exactly when a Type B event occurs and the call to $\textsc{WithProbability}$ in line 1 of \cref{alg:refresh} returns $\textsf{TRUE}$.
\end{itemize}

Thus, relevant events account for every possible recoloring of $v$. Moreover, each relevant event causes at most one direct invocation of $\textsc{Recolor}(v)$ and exactly one invocation of $\textsc{Refresh}(v)$. Within each relevant event, the direct recoloring stage, if it occurs, precedes the refresh stage. We therefore view each relevant event as consisting of two stages: (1) direct recoloring through $\textsc{Recolor}(v)$, followed by (2) probabilistic recoloring through $\textsc{Refresh}(v)$.

To prove \cref{lem:jump}, we need the following sublemma. Informally, suppose that a recoloring of \(v\) is triggered during the relevant event \(R_j\), and let \(L=k-j\) be the number of relevant events that occur afterwards.  The sublemma shows that, for any fixed randomization settled previously or from the lower levels, conditioned on no subsequent probabilistic recoloring of \(v\) being triggered, the probability that \(v\) ever receives the target color \(q^*\) after this recoloring is at most $O(\frac{L + \theta(v)}{\theta(v)^2})$. We will combine this with the fact that the probability that none of the next
\(L\) probabilistic recoloring opportunities for \(v\) is triggered is at most $\exp\left(-\Omega\left(\frac{L}{\theta(v)}\right)\right)$.

 \begin{lemma}\label{lem:jump_sublemma}
Fix the context of \cref{lem:jump}.  Let \(R_1,\ldots,R_k\) be the resulting
sequence of relevant events, and interpret \(R_0\) as the initial coloring stage
of \(v\).  In addition, fix the following:
\begin{itemize}
    \item an index \(j\in\{0,1,\ldots,k\}\);
    \item if \(j\ge 1\), either the direct recoloring stage of \(R_j\) or the
    probabilistic recoloring stage of \(R_j\), which we call the chosen
    recoloring stage; if \(j=0\), the chosen recoloring stage is the initial
    coloring stage of \(v\);
    \item all random choices made at vertices of level exactly \(l(v)\) before
    the chosen recoloring stage.  When \(j=0\), this means the initial colors of
    all vertices of level exactly \(l(v)\), except for \(v\) itself.
\end{itemize}

Condition on the event that the chosen recoloring stage triggers a recoloring of
\(v\), and that no subsequent probabilistic recoloring of \(v\) is
triggered afterwards.  Then the probability that \(c(v)=q^*\) at any point after
the recoloring of \(v\) during \(R_j\) is at most

\[
\frac{2(k-j)/\theta(v)+1}{\theta(v)}.
\]
\end{lemma}

\begin{proof}
We may assume that $k-j\le \theta(v)^2$, since otherwise the desired bound is trivial. Under this assumption, the condition $\Delta+1-d_{\leq}(v)\ge \theta(v)$ holds at every moment from $R_j$ and onward. To see this, first note that edge deletions cannot violate the condition: if the condition holds before deleting an edge, then it also holds afterward, since $d_{\leq}(v)$ can only decrease. Hence the last moment at which the condition could become violated is immediately after an edge insertion. By \cref{def:goodlevel}, and the fact that every edge insertion accounts for a Type A event, this occurs before $R_j$. Thus the condition holds throughout the interval we analyze.

Immediately after $v$ is recolored during the chosen recoloring stage of $R_j$, the probability that $c(v)=q^*$ is at most $1/\theta(v)$, as there are at least $\theta(v)$ colors that could be sampled.

Suppose that $c(v)\neq q^*$ immediately after this recoloring. Then $c(v)$ can become equal to $q^*$ only if $v$ is recolored again. Since we have conditioned on the event that none of the subsequent probabilistic recoloring of $v$ is triggered, any later recoloring of $v$ must be forced directly by one of the events $R_{j+1},R_{j+2},\ldots,R_k.$

We now identify a necessary condition for each such event to force the next recoloring of $v$.

First consider a Type A event $R_{j+t}$, where $t\ge 1$. Let the inserted edge be $(v,w)$. We claim that there is a color $q_{j+t}$, determined by the current context, such that $R_{j+t}$ can force the next recoloring of $v$ only if $c(v)=q_{j+t}$ immediately before $R_{j+t}$. Consider the three possible comparisons between $l(w)$ and $l(v)$:

\begin{itemize}
\item If $l(w)<l(v)$, then the conflicting color is simply the color of $w$ at the moment of the insertion. This color is determined by the fixed context.
\item If $l(w)=l(v)$, then the only possible conflicting color is the color of $w$ immediately before the chosen recoloring stage of $R_j$. Indeed, suppose that $v$ is forced to recolor at $R_{j+t}$ even though its color does not match the color that $w$ had immediately before the chosen recoloring stage of $R_j$. Then $w$ must have been recolored after the recoloring of $v$ during $R_j$. In that case, $t(w)>t(v)$, so the algorithm would recolor $w$, not $v$. Therefore this case cannot force a recoloring of $v$ and we reach a contradiction. Hence, when $l(w)=l(v)$, the relevant conflicting color is already determined before the recoloring of $v$ during $R_j$, and is therefore fixed by the current context.
\item If $l(w)>l(v)$, then the insertion cannot force $v$ to recolor, since the algorithm recolors toward the higher-level endpoint. Thus we can set $q_{j + t}$ arbitrarily.
\end{itemize}

Now consider a Type B event $R_{j+t}$, again with $t\ge 1$. The same conclusion holds: there is a color $q_{j+t}$, determined by the current context, such that $R_{j+t}$ can force the next recoloring of $v$ only if $c(v)=q_{j+t}$. In this case, $q_{j+t}$ is simply the new color of the lower-level neighbor of $v$ that has just been recolored, and this color is determined by the fixed context.

We prove the lemma by induction on the decreasing order of $j$. Order the colors in $\mathcal A_v\setminus\{q^*\}$ by the first time they appear among
\[
q_{j+1},q_{j+2},\ldots,q_k.
\]
Colors that never appear in this sequence may be placed arbitrarily after all colors that do appear.

Suppose the color sampled for $v$ is the $i$-th color in this order. We claim that in this case, the probability that $c(v)=q^*$ at some later point is at most
\[
\frac{2(k-j-i)/\theta(v)+1}{\theta(v)}
\]
when $j+i\le k$, and is $0$ otherwise. The induction hypothesis applies uniformly over all valid continuations of the random choices, as long as the corresponding event triggers a direct recoloring of $v$. Hence, we can condition on any randomization that had been chosen after $R_j$. Although we don't know what is the actual recoloring event that is triggered (if there is any), we know that it is no earlier than $R_{j + i}$, and among all valid possibilities the upper bound of $R_{j + i}$ dominates.

Let
\[
r=\min(\theta(v)-1,k-j).
\]
The total probability is at most
\[
\begin{aligned}
&\frac{1}{\theta(v)}
+
\sum_{i=1}^{r}
\frac{1}{\theta(v)}
\cdot
\frac{2(k-j-i)/\theta(v)+1}{\theta(v)}
\\
&=
\frac{1}{\theta(v)}
+
\frac{2}{\theta(v)^3}
\sum_{i=1}^{r}(k-j-i)
+
\frac{r}{\theta(v)^2}
\\
&=
\frac{1}{\theta(v)}
+
\frac{2r(k-j)}{\theta(v)^3}
-
\frac{r(r+1)}{\theta(v)^3}
+
\frac{r}{\theta(v)^2}
\\
&=
\frac{1}{\theta(v)}
+
\frac{2r(k-j)+r(\theta(v)-r-1)}{\theta(v)^3}.
\end{aligned}
\]

It remains to bound the numerator in the second term. If $r=k-j$, then
\[
2r(k-j)+r(\theta(v)-r-1)
=
2r^2+r(\theta(v)-r-1)
\le
2\theta(v)r
=
2\theta(v)(k-j).
\]
If $r=\theta(v)-1$, then
\[
2r(k-j)+r(\theta(v)-r-1)
=
2(\theta(v)-1)(k-j)
\le
2\theta(v)(k-j).
\]
Therefore,
\[
\begin{aligned}
\frac{1}{\theta(v)}
+
\frac{2r(k-j)+r(\theta(v)-r-1)}{\theta(v)^3}
&\le
\frac{1}{\theta(v)}
+
\frac{2\theta(v)(k-j)}{\theta(v)^3}
\\
&=
\frac{2(k-j)/\theta(v)+1}{\theta(v)}.
\end{aligned}
\]
This proves the lemma.
\end{proof}

To prove \cref{lem:jump2}, we also show the analogue of \cref{lem:jump_sublemma}:

\begin{lemma}\label{lem:jump2_sublemma}
Fix the context of \cref{lem:jump2}.  Let \(R_1,\ldots,R_k\) be the resulting
sequence of relevant events, and interpret \(R_0\) as the initial coloring stage
of \(v\).  In addition, fix the following:
\begin{itemize}
    \item an index \(j\in\{0,1,\ldots,k\}\);
    \item if \(j\ge 1\), either the direct recoloring stage of \(R_j\) or the
    probabilistic recoloring stage of \(R_j\), which we call the chosen
    recoloring stage; if \(j=0\), the chosen recoloring stage is the initial
    coloring stage of \(v\);
    \item all random choices made at vertices of level exactly \(l(v)\) before
    the chosen recoloring stage.  When \(j=0\), this means the initial colors of
    all vertices of level exactly \(l(v)\), except for \(v\) itself.
\end{itemize}

Condition on the event that the chosen recoloring stage triggers a recoloring of
\(v\), and that no subsequent probabilistic recoloring of \(v\) is
triggered afterwards.  Then the probability that $c(u)=c(v)$ and $t(u)<t(v)$ at any point after the recoloring of $v$ during $R_j$ is at most
\[
\frac{2(k-j)/\theta(v)+1}{\theta(v)}.
\]
\end{lemma}

\begin{proof}
The difference between \cref{lem:jump_sublemma} and the present lemma is that we now bound the event
\[
c(u)=c(v) \quad\text{and}\quad t(u)<t(v)
\]
where $u$ is a vertex at the same level as $v$, instead of bounding the event $c(v)=q^*$ for a fixed color $q^*$.

We may assume that $k-j\le \theta(v)^2$, since otherwise the desired bound is trivial. Under this assumption, the condition $\Delta+1-d_{\leq}(v)\ge \theta(v)$ holds at every moment from $R_j$ and onward. To see this, first note that edge deletions cannot violate the condition: if the condition holds before deleting an edge, then it also holds afterward, since $d_{\leq}(v)$ can only decrease. Hence the last moment at which the condition could become violated is immediately after an edge insertion. By \cref{def:goodlevel}, and the fact that every edge insertion accounts for a Type A event, this occurs before $R_j$. Thus the condition holds throughout the interval we analyze.

Immediately after $v$ is recolored during the chosen recoloring stage of $R_j$, the probability that $c(v)=c(u)$ is at most $1/\theta(v)$, as there are at least $\theta(v)$ colors that could be sampled. Here $c(u)$ is fixed by the current context. Since we only need an upper bound, we do not need to use the condition $t(u)<t(v)$ if this case happens.

Suppose that $c(v)\neq c(u)$ immediately after this recoloring. Then the event $c(u)=c(v)$ and $t(u)<t(v)$ can occur only after $v$ is recolored again. Indeed, recoloring $u$ may make $c(u)=c(v)$ true, but if this happens without a later recoloring of $v$, then the timestamp of $u$ becomes larger than the timestamp of $v$, violating the condition $t(u)<t(v)$. Since we have conditioned on the event that none of the subsequent probabilistic recoloring of $v$ is triggered, any later recoloring of $v$ must be forced directly by one of the events $R_{j+1},R_{j+2},\ldots,R_k$.

By identifying a necessary condition under which each such event can force the
next recoloring of \(v\), we reduce to exactly the setting considered in the
proof of \cref{lem:jump_sublemma}.  The same inductive argument then yields the
upper bound $\frac{2(k-j)/\theta(v)+1}{\theta(v)}$.
\end{proof}

We now prove \cref{lem:jump} and \cref{lem:jump2}.

\begin{proof}[Proof of \cref{lem:jump}]
Let $J$ be the index of the last relevant event that triggers a probabilistic recoloring of $v$. If no relevant event triggers a probabilistic recoloring of $v$, set $J=0$.

We bound
\[
\sum_{j=0}^{k}
\Pr[c(v)=q^* \mid J=j]\cdot \Pr[J=j].
\]
By \cref{lem:jump_sublemma}, this is at most
\[
\frac{2k/\theta(v)+1}{\theta(v)}
\left(1-\frac{10^{-4}}{\theta(v)}\right)^k
+
\sum_{j=1}^{k}
\frac{2(k-j)/\theta(v)+1}{\theta(v)}
\cdot
\frac{10^{-4}}{\theta(v)}
\left(1-\frac{10^{-4}}{\theta(v)}\right)^{k-j}.
\]

It remains to bound this expression. Let
\[
c=10^{-4}
\qquad\text{and}\qquad
q=1-\frac{c}{\theta(v)}.
\]
Then the expression becomes
\[
\begin{aligned}
&\frac{2k/\theta(v)+1}{\theta(v)}q^k
+
\sum_{j=1}^{k}
\frac{2(k-j)/\theta(v)+1}{\theta(v)}
\cdot
\frac{c}{\theta(v)}
\cdot
q^{k-j}
\\
&=
\frac{2k/\theta(v)+1}{\theta(v)}q^k
+
\frac{c}{\theta(v)}
\sum_{t=0}^{k-1}
\frac{2t/\theta(v)+1}{\theta(v)}q^t .
\end{aligned}
\]

For the first term, using $q^k\le \exp(-ck/\theta(v))$, we get
\[
\begin{aligned}
\frac{2k/\theta(v)+1}{\theta(v)}q^k
&\le
\frac{1}{\theta(v)}
\left(\frac{2k}{\theta(v)}+1\right)
\exp\left(-\frac{ck}{\theta(v)}\right)
\\
&\le
\frac{1}{\theta(v)}
\left(\frac{2}{ec}+1\right).
\end{aligned}
\]

For the second term, we have
\[
\begin{aligned}
\frac{c}{\theta(v)}
\sum_{t=0}^{k-1}
\frac{2t/\theta(v)+1}{\theta(v)}q^t
&\le
\frac{c}{\theta(v)}
\sum_{t=0}^{\infty}
\frac{2t/\theta(v)+1}{\theta(v)}q^t
\\
&=
\frac{c}{\theta(v)}
\left(
\frac{1}{\theta(v)}\sum_{t=0}^{\infty}q^t
+
\frac{2}{\theta(v)^2}\sum_{t=0}^{\infty}tq^t
\right)
\\
&=
\frac{c}{\theta(v)}
\left(
\frac{1}{\theta(v)}\cdot \frac{1}{1-q}
+
\frac{2}{\theta(v)^2}\cdot \frac{q}{(1-q)^2}
\right)
\\
&=
\frac{c}{\theta(v)}
\left(
\frac{1}{c}
+
\frac{2q}{c^2}
\right)
\\
&\le
\frac{1}{\theta(v)}
+
\frac{2}{c\theta(v)}.
\end{aligned}
\]

Combining the two bounds gives
\[
\frac{2k/\theta(v)+1}{\theta(v)}q^k
+
\sum_{j=1}^{k}
\frac{2(k-j)/\theta(v)+1}{\theta(v)}
\cdot
\frac{c}{\theta(v)}
\cdot
q^{k-j}
=
O\left(\frac{1}{\theta(v)}\right).
\]
This completes the proof.
\end{proof}

\begin{proof}[Proof of \cref{lem:jump2}]
Let $J$ be the index of the last relevant event that triggers a probabilistic recoloring of $v$. If no relevant event triggers a probabilistic recoloring of $v$, set $J=0$.

We bound
\[
\sum_{j=0}^{k}
\Pr[c(u)=c(v)\land t(u)<t(v) \mid J=j]\cdot \Pr[J=j].
\]
By \cref{lem:jump2_sublemma}, this is at most
\[
\frac{2k/\theta(v)+1}{\theta(v)}
\left(1-\frac{10^{-4}}{\theta(v)}\right)^k
+
\sum_{j=1}^{k}
\frac{2(k-j)/\theta(v)+1}{\theta(v)}
\cdot
\frac{10^{-4}}{\theta(v)}
\left(1-\frac{10^{-4}}{\theta(v)}\right)^{k-j}.
\]

As shown in the proof of \cref{lem:jump}, this expression is upper bounded by $O(\frac{1}{\theta(v)})$.
\end{proof}

\begin{lemma}\label{lem:EVcollideC}
For any edge insertion \((u,v)\) and for any level assignment $l$ that is good for \(v\), the
probability that
\[
(l(u),t(u))<_{\mathrm{lex}}(l(v),t(v))
\qquad\text{and}\qquad
c(u)=c(v)
\]
immediately before the insertion is \(O(1/\theta(v))\).
\end{lemma}

\begin{proof}
Fix the graph and the prefix of the update sequence ending immediately
before the insertion of \((u,v)\). Consider the execution of the data
structure on this fixed prefix, conditioned on the level assignment being $l$. Fix all
random choices made at vertices whose level is less than \(l(v)\).

If \(l(u)<l(v)\), by \cref{obs:hierarchy}, the color \(c(u)\)
is fixed under the current conditioning; denote it by \(q^*\). Since \(l\)
is good for \(v\), \cref{lem:jump} implies that, over the remaining
randomness, $\Pr[c(v)=q^*]=O(1/\theta(v))$. 
Thus, in this case, the probability that
\((l(u),t(u))<_{\mathrm{lex}}(l(v),t(v))\) and \(c(u)=c(v)\) is
\(O(1/\theta(v))\).

If \(l(u)=l(v)\), since \(l\) is good for \(v\),
\cref{lem:jump2} gives $\Pr[t(u)<t(v)\land c(u)=c(v)]=O(1/\theta(v))$. 

Therefore, for every fixed outcome of the random choices made below level \(l(v)\),
the probability in the statement is \(O(1/\theta(v))\). Averaging over the random choices made below level \(l(v)\)
preserves this bound.
\end{proof}

\subsection{Expected Time For Each Recoloring}\label{sec:amort-analysis-cost}
Here, we present an upper bound on the expected time that the procedure $\textsc{Recolor}(v)$ takes when it is triggered, including all the other $\textsc{Recolor}(w)$ procedures that it causes to be triggered, directly or indirectly. Let us clarify this: when the procedure $\textsc{Recolor}(v)$ is invoked, besides coloring node $v$, we may invoke $\textsc{Recolor}(w)$ for a number of nodes $w\in NB_{>}(v)$, either because their prior color is equal to the new color that $v$ selected, or simply because of the $\textsc{Refresh}(w)$ called on $w$ which recolors is with some (small) probability. We call this a direct trigger of $w$ by $v$. Furthermore, each such $w$ may further directly trigger other $w'\in B_{>}(w)$ to be recolored similarly, and this can continue iteratively. In such a case, we say $w'$ is indirectly triggered by the recoloring of $v$. Assuming $v$ is in level $l(v)$, the process can continue at most $\log \Delta-l(v)$ iterations, since after that there is no neighbor that can be triggered. Below, we bound the expected time for processing all the $\textsc{Recolor}()$ procedures triggered, directly or indirectly, starting from the procedure $\textsc{Recolor}(v)$ for a node $v$ in level $l(v)$.

Fix a vertex $v$ and its level $l(v)$, and let $d=\Delta\cdot2^{-l(v)}$. We call a vertex $u\neq v$ \textbf{relevant} for $v$ if $l(u)>l(v)$, and let
\[
M_v=\{v\}\cup\{u:l(u)>l(v)\}.
\]
If $v$ is clear from context, we simply call $u$ relevant and write $M$ for $M_v$. If $l(v)<\lceil\log\Delta\rceil+1$, each vertex $u\neq v$ is relevant for $v$ independently with probability $2^{-l(v)}$. At the maximum level, $M_v=\{v\}$. Let $U$ be the set of vertices at distance at most $\log d+2$ from $v$ in $G[M_v]$. We evaluate the graph and all neighborhoods below immediately after the current edge insertion and before processing its recolorings.

\begin{definition}\label{def:typicallevel}
Fix sufficiently large absolute constants $C_0$ and $C$, and let $\hat d=C_0d$. We call a level assignment \textbf{typical} for $v$ if it satisfies the following two conditions:
\begin{enumerate}
    \item Every vertex $w\in U$ has $\deg_{G[M_v]}(w)\leq\hat d$, and $|NB_{\geq}(v)|\leq\hat d$.
    \item For every vertex $w\in U$ and every integer $j\geq0$, the number of neighbors of $w$ in level $l(v)+j$ is at most
    \[
    4\frac{d}{2^j}+C\log^2d.
    \]
\end{enumerate}
Otherwise, we call the level assignment \textbf{atypical} for $v$.
\end{definition}

Let $L$ denote the level assignment on $V\setminus\{v\}$, let $R$ denote the random choices made before the current update, and let $R'$ denote the fresh random choices used to process the current update. Let $\mathcal W_v$ denote the running time of a potential invocation of $\textsc{Recolor}(v)$, including all recursive calls it triggers. For the initial conflict-triggered invocation, its initial state is determined by $L$ and $R$. We also prove the same bound for invocations later in the update, conditional on the choices from $R'$ preceding them.

\begin{lemma}\label{lem:EVrecolor}
    Fix a vertex $v$ and its level $l(v)$. The expected running time of $\textsc{Recolor}(v)$, including all recursive calls it triggers, is
    \[
    \E_{L,R,R'}[\mathcal W_v]=O(\Delta\cdot2^{-l(v)})=O(\theta(v)).
    \]
    More precisely, let $\mathcal T_v$ be the event that the level assignment is typical for $v$, as in \cref{def:typicallevel}. Then:
    \begin{enumerate}
        \item For every typical assignment $L=\ell$ and every fixed choice $R=r$ of the randomness before the current update,
        \[
        \E_{R'}[\mathcal W_v\mid L=\ell,R=r]=O(\theta(v)).
        \]
        \item The contribution from atypical assignments satisfies
        \[
        \E_{L,R,R'}[\mathbf 1[\lnot\mathcal T_v]\mathcal W_v]=O(1).
        \]
    \end{enumerate}
    The bound in item 1 holds uniformly over the initial state of the recoloring. Moreover, the bound $\E_{L,R,R'}[\mathcal W_v]=O(\theta(v))$ remains valid after conditioning on the event that the level assignment is bad for $v$, as defined in \cref{def:goodlevel}.
\end{lemma}

To prove \Cref{lem:EVrecolor}, we first bound the degrees near $v$ in $G[M_v]$, using the set $M_v$ defined before \cref{def:typicallevel}. We take a union bound over short paths from $v$. Requiring all vertices of a path to belong to $M_v$ offsets the number of possible paths, while conditioning on their inclusion leaves all other membership choices independent.

\begin{lemma}\label{lem:climbbound_claim1}
    Fix a vertex $v$ and its level $l(v)$, and let $d=\Delta\cdot2^{-l(v)}$. For a sufficiently large absolute constant $c_0$ and every $X\geq c_0d$, with probability at least $1-\exp(-\Omega(X))$, every vertex $w$ at distance at most $\log d+2$ from $v$ in $G[M_v]$ has $\deg_{G[M_v]}(w)\leq2X$. The probability is over the randomness of $M_v$.
\end{lemma}
\begin{proof}
If $l(v)=\lceil\log\Delta\rceil+1$, then $M_v=\{v\}$ and the claim is immediate. Otherwise, each vertex other than $v$ belongs to $M_v$ independently with probability $p=2^{-l(v)}=d/\Delta$. Let $h=\lfloor\log d+2\rfloor$. If a vertex $w$ violates the desired degree bound, there is a simple path $P=(v=v_0,v_1,\ldots,v_r=w)$ of length $r\leq h$ whose vertices all belong to $M_v$.

Fix such a path in $G$. All its vertices belong to $M_v$ with probability $p^r$. Conditional on their inclusion, the membership choices outside the path remain independent. At most $r$ neighbors of $w$ lie on the path, and the number of its remaining neighbors in $M_v$ is a sum of independent Bernoulli variables with expectation at most $\Delta p=d$. Choose $c_0$ large enough that $X\geq c_0d$ implies $X\geq h$ and $X\geq2d$. Chernoff's bound gives
\[
\Pr\!\left[\deg_{G[M_v]}(w)>2X\,\middle|\,V(P)\subseteq M_v\right]
\leq\exp(-c_1X)
\]
for an absolute constant $c_1>0$. There are at most $\Delta^r$ simple paths of length $r$ starting at $v$. Taking a union bound over these paths and their lengths, we obtain
\[
\begin{aligned}
\Pr\!\left[\exists w\in U:\deg_{G[M_v]}(w)>2X\right]
&\leq\sum_{r=0}^{h}\Delta^rp^r\exp(-c_1X)\\
&=\left(\sum_{r=0}^{h}d^r\right)\exp(-c_1X)\\
&\leq\exp\!\left(O(\log^2(d+2))-c_1X\right)\\
&\leq\exp(-\Omega(X)).
\end{aligned}
\]
The last inequality follows by choosing $c_0$ sufficiently large. This also covers bounded $d$, since the level assignment guarantees $d>1/4$.
\end{proof}

\begin{lemma}\label{lem:climbbound_claim3}
\cref{lem:climbbound_claim1} is true, even when conditioned on the level assignment that is bad for $v$. Notice that, to see if the level assignment is good or bad, it suffice to reveal for each such node $u\in V - \{v\}$ only whether $u\in M_{v}$ or not.
\end{lemma}

\begin{proof}
Let \(\mathcal E_1\) be the event in \cref{lem:climbbound_claim1}, and let
\(\mathcal E_2\) be the event that the leveling scheme is bad. We view both
events as functions of the indicator vector $\bigl(\mathbf 1[v_1\in M_v],\mathbf 1[v_2\in M_v],\ldots,
\mathbf 1[v_{|V|-1}\in M_v]\bigr)$ 
where \(\{v_1,v_2,\ldots,v_{|V|-1}\}=V\setminus\{v\}\).

We first show that \(\mathcal E_1\) is decreasing with respect to this
indicator vector. Suppose that \(\mathcal E_1\) fails when \(M_v=S\). Then
there exists a path of length at most \(\log d+2\) from \(v\) to a vertex
\(w\) such that the path is contained in \(S\) and
\(\deg_{G[S]}(w)>2X\). For any \(T\supseteq S\), the same path is still
contained in \(T\), and $\deg_{G[T]}(w)\geq \deg_{G[S]}(w)>2X$.
Hence \(\mathcal E_1\) also fails when \(M_v=T\). Therefore, \(\mathcal E_1\) is
decreasing.

We next show that \(\mathcal E_2\) is also decreasing. Suppose that
\(\mathcal E_2\) fails when \(M_v=S\); equivalently, the level assignment
is good for \(v\) with respect to \(S\). If we replace \(S\) by any superset
\(T\supseteq S\), then \(d_{\leq}(v)\) can only decrease. Consequently,
\(\Delta+1-d_{\leq}(v)\) can only increase. Thus every inequality of the
form $\Delta+1-d_{\leq}(v)\geq \theta(v)$
that held for the last \(2\theta(v)^2\) relevant neighborhoods continues to
hold after replacing \(S\) by \(T\). Hence the level assignment remains good
for \(v\), so \(\mathcal E_2\) fails. Therefore, \(\mathcal E_2\) is decreasing.

Since both \(\mathcal E_1\) and \(\mathcal E_2\) are decreasing events,
Harris' inequality (\cref{thm:harris}) gives $\Pr[\mathcal E_1\mid \mathcal E_2]\geq \Pr[\mathcal E_1]$. 
Thus the bound from \cref{lem:climbbound_claim1} continues to hold after
conditioning on \(\mathcal E_2\).
\end{proof}

Now we are ready to present the proof of \cref{lem:EVrecolor}. 


\begin{proof}[Proof of \cref{lem:EVrecolor}]
Let $d=\Delta\cdot2^{-l(v)}$ and $\hat d=C_0d$, as in \cref{def:typicallevel}. We call each neighbor $w\in NB_{>}(v)$ an out-neighbor of $v$. The procedure $\textsc{Recolor}(v)$ itself takes $O(1+|NB_{\geq}(v)|)$ expected time to sample the color of $v$ and update the relevant data structures. It then scans all out-neighbors and may trigger their recolorings, which may trigger further recolorings. We bound the expected time for all these calls. Each recursive call moves to a strictly higher level, so every chain has at most $\lceil\log\Delta\rceil+1-l(v)\leq\log d+2$ edges and stays within the set $U$ defined before \cref{def:typicallevel}.

We first bound the contribution from atypical assignments. We begin with the first condition of \cref{def:typicallevel} and consider the maximum of $|NB_{\geq}(v)|$ and the degrees $\deg_{G[M_v]}(w)$ over $w\in U$. Including $|NB_{\geq}(v)|$ accounts for the same-level neighbors scanned by the initial call. For every other vertex reached by the cascade, all of its same-level and higher-level neighbors already belong to $M_v$. By \cref{lem:climbbound_claim1} and Chernoff's bound for $|NB_{\geq}(v)|$, whose expectation is at most $2d$, the probability that this maximum lies in $[X,2X]$ is at most $\exp(-\Omega(X))$ for $X\geq\hat d$, provided $C_0$ is sufficiently large. We bound the cost in this case by allowing each call to trigger all of its out-neighbors through both collisions and refreshes. This gives at most $(O(X))^{O(\log d+2)}\leq\exp(O(\log^2 X))$ operations on the data structures, each of constant expected cost. We obtain this bound for every initial state. Hence, the contribution from assignments that violate the first condition is at most
\[
\sum_{X\geq\hat d}\exp(-\Omega(X))\exp(O(\log^2 X))=O(1).
\]
If $d$ is bounded, the first condition bounds both the local degrees and the recursion depth by constants. Thus, atypical assignments satisfying this condition contribute $O(1)$, and every typical assignment has $O(1)$ expected cost. We therefore assume below that $d$ is sufficiently large.

Next, we bound the probability of violating the second condition. We use the same path argument as in \cref{lem:climbbound_claim1}. Fix a simple path $P$ of length $r\leq h=\lfloor\log d+2\rfloor$ from $v$ to $w$, and condition on $V(P)\subseteq M_v$. For each level $l(v)+j$, where $j\geq0$, the number of neighbors of $w$ outside the path in that level is a sum of independent Bernoulli variables with expectation at most $2d/2^j$. This bound includes the doubled probability mass at the maximum level. The path contributes at most $r$ additional neighbors. Since $r=O(\log d)$, Chernoff's bound gives
\[
\Pr\!\left[\#\{u:\{u,w\}\in E,\ l(u)=l(v)+j\}>4d/2^j+C\log^2d
\,\middle|\,V(P)\subseteq M_v\right]
\leq\exp(-\Omega(C\log^2d)).
\]
Taking a union bound over the paths and the $O(\log d)$ possible levels, and using $\Pr[V(P)\subseteq M_v]=p^r$ for $p=d/\Delta$, we bound the probability of violating the second condition by
\[
O(\log d)\left(\sum_{r=0}^{h}d^r\right)\exp(-\Omega(C\log^2d))
=\exp(-\Omega(C\log^2d)),
\]
for sufficiently large $C$. If the first condition holds, the number of operations is at most $\hat d^{O(\log d)}$, even if every call triggers all of its out-neighbors through both collisions and refreshes. Therefore, assignments that satisfy the first condition but violate the second contribute at most
\[
\exp(-\Omega(C\log^2d))\hat d^{O(\log d)}=O(1)
\]
to the expected cost. Combining the two exceptional-case estimates gives
\[
\E_{L,R,R'}[\mathbf 1[\lnot\mathcal T_v]\mathcal W_v]=O(1).
\]
We used cost bounds uniform over the initial state, so these estimates also apply to the state produced by the preceding updates.

For the remainder of the unconditioned analysis, fix any typical assignment $L=\ell$, as in \cref{def:typicallevel}, and any past randomness $R=r$. We take all expectations below over the fresh randomness $R'$ used by the recoloring and its recursive calls. We prove the bounds uniformly over the initial state and over the execution history preceding each recursive invocation.

We now represent the recursive calls by a token-spreading process. Initially, a single token resides at $v$. A token at a vertex $u$ in level $l(v)+i$ generates new tokens at its out-neighbors through two processes. First, the new color of $u$ may conflict with the colors of its out-neighbors. If $u$ has $t$ out-neighbors, it samples from at least $t+1$ available colors, so the expected number of these conflicts is at most $t/(t+1)<1$. Subsequent recursive calls cannot create new conflicts with $u$, since they avoid the colors of their lower-level neighbors. Second, each out-neighbor $w$ in level $l(v)+j$, where $j>i$, receives a token through a refresh with probability
\[
\frac{10^{-4}}{\theta(w)}
=\frac{10^{-4}}{\lceil d2^{-j}/100\rceil}
\leq\frac{1}{100}\frac{2^j}{d}.
\]
After these two processes, the token at $u$ terminates. Each token incurs $O(1+|NB_{\geq}(u)|)$ expected local cost. We bound the expected total cost, counting repeated visits to a vertex with multiplicity.

Let $j^*=\lceil\log_2(d/(100C\log^2d))\rceil$. We call a level $l(v)+j$ \textit{low} if $j<j^*$ and \textit{high} otherwise. For sufficiently large $d$, we have $0<j^*<\log d$ and $j^*\geq\log d-O(\log\log d)$. Typicality bounds the number of neighbors in each low level $l(v)+j$ by $4d/2^j+C\log^2d<4.01d/2^j$. There are only $O(\log\log d)$ high levels before the maximum level $\lceil\log\Delta\rceil+1$, and each contributes $O(\log^2d)$ neighbors. Thus, every token in a high level scans at most $O(\log^3d)$ same-level and higher-level neighbors. Even if each call triggers every out-neighbor through both processes, the number of calls and data-structure operations starting from a high-level token is at most $(O(\log^3d))^{O(\log\log d)}$. Choose
\[
K=(O(\log^3d))^{O(\log\log d)}
\]
sufficiently large to bound the expected cost of any such token, including all the calls it triggers.

We define $Q(j^*)=K$ and, for $0\leq i<j^*$,
\[
Q(i)=a\frac{d}{2^i}+Q(i+1)
+\frac{5}{100}\sum_{j=i+1}^{j^*}Q(j)
+a\log^3d\,K,
\]
where $a$ is a sufficiently large absolute constant. We prove by reverse induction on $i$ that the initial token at $v$ and every token descended from it satisfy the following bound: if the token is at a vertex in level $l(v)+i$, its expected remaining cost is at most $Q(i)$, conditional on the history preceding its processing. The high-level bound proves the base case. Notice also that $Q(i)\geq Q(i+1)\geq K$.

For the induction step, consider a token at $u$ in level $l(v)+i$, where $i<j^*$. Summing the neighbor bounds over the low and high levels gives $O(d/2^i+\log^3d)$ expected local cost. The collision process generates at most one token in expectation, and each such token has expected cost at most $Q(i+1)$ by induction and the high-level bound. For the refresh process, each low level $l(v)+j$ with $i<j<j^*$ receives at most
\[
\frac{4.01d}{2^j}\cdot\frac{1}{100}\frac{2^j}{d}
\leq\frac{5}{100}
\]
tokens in expectation, each with expected cost at most $Q(j)$. We apply the induction hypothesis conditional on the history preceding each child invocation; it holds for every such history. There are at most $O(\log^3d)$ neighbors in high levels, each of which can receive one refresh token with expected cost at most $K$. The last term of the recurrence absorbs both these high-level contributions and the $O(\log^3d)$ part of the local scan. By linearity of expectation, the total expected cost is at most $Q(i)$.

We next bound this recurrence. Reverse induction gives
\[
Q(i)\leq A\frac{d}{2^i}
+B\sqrt{\frac{d}{2^i}}\,\log^3d\,K
\]
for sufficiently large absolute constants $A$ and $B$. Indeed, substituting this bound for the later terms of the recurrence multiplies the two terms by factors at most
\[
\frac12+\frac{5}{100}<1
\qquad\text{and}\qquad
\frac{1}{\sqrt2}+\frac{5}{100}\frac{1}{\sqrt2-1}<1,
\]
respectively. We absorb the remaining terms by choosing $A$ and $B$ sufficiently large; the same choices cover the base case. Since $\log^3d\,K=d^{o(1)}$, we obtain $Q(0)=O(d)$. Consequently, for every typical assignment and every fixed past randomness,
\[
\E_{R'}[\mathcal W_v\mid L=\ell,R=r]\leq Q(0)=O(d)=O(\theta(v)).
\]
Combining this bound with the atypical contribution proves the claimed expected running time.

Finally, we prove the bound conditioned on the assignment being bad for $v$, as in \cref{def:goodlevel}, whenever that event has positive probability. The bad-level event depends only on $M_v$. By \cref{lem:climbbound_claim3}, the local degree tail bound continues to hold under this conditioning. The number of higher-level neighbors of $v$ is increasing in $M_v$, so Harris' inequality also bounds its conditional upper tail by its unconditional upper tail. For the same-level neighbors, fix $M_v$ and reveal the remaining levels. Below the maximum level, each vertex outside $M_v$ has level $l(v)$ with conditional probability $2^{-l(v)}/(1-2^{-l(v)})\leq2\cdot2^{-l(v)}$; at the maximum level, this probability is $2^{1-l(v)}$. These choices remain independent after conditioning on $M_v$ and on the assignment being bad. Their expected contribution is at most $2d$. Thus, the same tail and cost estimates bound the contribution from violations of the first condition by $O(1)$. For bounded $d$, the remaining cost is $O(1)$ as before.

For sufficiently large $d$, we also repeat the second-condition estimate. For levels $l(v)+j$ with $j\geq1$, we can generate the assignment by independently sampling membership in $M_v$ and, for every vertex other than $v$, an independent residual level to use if it belongs to $M_v$. The residual levels follow the geometric distribution with its remaining mass combined at the maximum level. For every fixed choice of these residual levels, the event that some $w\in U$ violates a higher-level neighbor bound is increasing in the membership indicators of $M_v$. The bad-level event is decreasing, so Harris' inequality shows that conditioning on it cannot increase this violation probability. Averaging over the residual levels preserves the bound $\exp(-\Omega(C\log^2d))$.

For $j=0$, fix any $M_v$ satisfying the local degree bound. Then $|U|\leq\hat d^{O(\log d)}$, and the level-$l(v)$ neighbors of each $w\in U$, apart from the possible neighbor $v$, are independent choices outside $M_v$ with total expectation at most $2d$. The bad-level event does not change these choices. Chernoff's bound and a union bound over $U$ therefore give the same $\exp(-\Omega(C\log^2d))$ estimate for violating the level-$l(v)$ bound. Combining these estimates with the coarse cost bound gives an $O(1)$ conditional atypical contribution. The uniform bound for typical assignments then yields
\[
\E_{L,R,R'}[\mathcal W_v\mid\lnot\mathcal G_v]=O(\theta(v)),
\]
where $\mathcal G_v$ denotes the event that the assignment is good for $v$.
\par
\end{proof}

\subsection{Proof of \cref{thm:sequential-main}}

We are now ready to prove the main theorem of this section.

\seqA*
\begin{proof}[Proof of \cref{thm:sequential-main}]
The initialization takes $O(n)$ time by initializing the hash sets of $NB_*$ and $\textsc{LowerEqualUsed}$ in $O(1)$ time for each vertex $v \in V$.

The expected running time of lines~1--3 of \cref{alg:insertedge} and of the entire procedure \cref{alg:deleteedge} is $O(1)$. We next bound the contribution of the two calls to $\textsc{Refresh}$ at the end of \cref{alg:insertedge}. Fix one endpoint $w$ and its level $l(w)$. We use $L$, $R$, and $R'$ as defined before \cref{lem:EVrecolor}, with $w$ as the distinguished vertex. For a typical assignment, we condition on $L$, $R$, and the choices from $R'$ preceding this refresh, including any earlier recolorings during the current update. The refresh triggers with probability $10^{-4}/\theta(w)$. By the uniform bound in \cref{lem:EVrecolor}, the expected cost whenever it triggers, over the unused choices from $R'$, is $O(\theta(w))$. Thus, for every typical $\ell$ and every $r$,
\[
\E_{R'}[\text{cost attributed to this refresh}\mid L=\ell,R=r]=O(1).
\]
For atypical assignments, we use the coarse cost bounds from the proof of \cref{lem:EVrecolor}. These bounds hold uniformly over the state preceding the refresh and give an $O(1)$ contribution to the expectation over $L,R,R'$. Averaging over $l(w)$ and adding the contributions of both endpoints gives $O(1)$ expected time. It remains to bound the cost of the recoloring triggered by a same-color conflict.
\par

Consider a fixed insertion of an unordered edge \(\{u,v\}\), where \(u\)
and \(v\) denote the two endpoints before the possible reordering in
lines~2--3 of \cref{alg:insertedge}. Since this reordering depends on the
random pairs \((l(u),t(u))\) and \((l(v),t(v))\), the endpoint that appears
second after the reordering is itself random. To avoid this dependency, we
bound separately the expected cost of a possible invocation of
\(\textsc{Recolor}(v)\); the same argument applies symmetrically to
\(\textsc{Recolor}(u)\).

Fix a possible value \(l'\) of \(l(v)\), and condition on the event
\(l(v)=l'\). All probabilities and expectations below are taken in this
conditional probability space. In particular, \(\theta(v)\) is fixed
throughout the remainder of the argument.

Let \(\mathcal P_v\) be the event that, immediately before the
current insertion,
\[
(l(u),t(u))<_{\mathrm{lex}}(l(v),t(v))
\qquad\text{and}\qquad
c(u)=c(v).
\]
Let $I_v=\mathbf 1[\mathcal P_v]$ be the indicator of $\mathcal P_v$. The event \(\mathcal P_v\) is exactly the condition under
which \(\textsc{Recolor}(v)\) is invoked in line~5.

Recall that $\mathcal W_v$ denotes the running time of a potential invocation of $\textsc{Recolor}(v)$, including all recursive calls it triggers. We use $L$, $R$, and $R'$ for the level assignment, the past randomness, and the fresh randomness, respectively, as in \cref{lem:EVrecolor}.
Up
to an \(O(1)\) additive term, the contribution of a possible invocation of
\(\textsc{Recolor}(v)\) is \(I_v\mathcal W_v\). It therefore suffices to
prove that
\[
\E_{L,R,R'}[I_v\mathcal W_v]=O(1).
\]

Let $\mathcal G_v$ be the event that the assignment is good for $v$, as in \cref{def:goodlevel}, and let $\mathcal T_v$ be the event that it is typical for $v$, as in \cref{def:typicallevel}. Both events are determined solely by $L$. We split the expected cost into three contributions:
\[
\begin{aligned}
\E_{L,R,R'}[I_v\mathcal W_v]
={}&\E_{L,R,R'}[\mathbf 1[\mathcal G_v\cap\mathcal T_v]I_v\mathcal W_v]\\
&+\E_{L,R,R'}[\mathbf 1[\lnot\mathcal T_v]I_v\mathcal W_v]\\
&+\E_{L,R,R'}[\mathbf 1[\lnot\mathcal G_v\cap\mathcal T_v]I_v\mathcal W_v].
\end{aligned}
\]

First, fix an assignment $L=\ell$ that is both good and typical. Once we fix $L=\ell$ and $R=r$, we determine $I_v$. By \cref{lem:EVrecolor}, we can bound the expected future cost by $O(\theta(v))$ for every $r$. Therefore,
\[
\begin{aligned}
\E_{R,R'}[I_v\mathcal W_v\mid L=\ell]
&=\E_R\!\left[I_v\,\E_{R'}[\mathcal W_v\mid L=\ell,R]\,\middle|\,L=\ell\right]\\
&\le O(\theta(v))\Pr_R[\mathcal P_v\mid L=\ell]\\
&=O(1),
\end{aligned}
\]
where the last equality follows from \cref{lem:EVcollideC}. Averaging over assignments that are both good and typical bounds the first contribution by $O(1)$.

Second, since $0\le I_v\le1$, we use \cref{lem:EVrecolor} to bound the atypical contribution by
\[
\E_{L,R,R'}[\mathbf 1[\lnot\mathcal T_v]I_v\mathcal W_v]
\le\E_{L,R,R'}[\mathbf 1[\lnot\mathcal T_v]\mathcal W_v]=O(1).
\]

Finally, consider assignments that are typical but not good. We apply Chernoff's bound from \cref{thm:chernoffva} at each of the last $2\theta(v)^2$ relevant times and take a union bound to obtain
\[
\Pr_L[\lnot\mathcal G_v]\le2\theta(v)^2\exp(-\Omega(\theta(v))).
\]
The uniform expected-cost bound in \cref{lem:EVrecolor} applies to every typical assignment, whether or not it is good. Thus,
\[
\begin{aligned}
\E_{L,R,R'}[\mathbf 1[\lnot\mathcal G_v\cap\mathcal T_v]I_v\mathcal W_v]
&\le O(\theta(v))\Pr_L[\lnot\mathcal G_v\cap\mathcal T_v]\\
&\le O\!\left(\theta(v)^3\exp(-\Omega(\theta(v)))\right)\\
&=O(1).
\end{aligned}
\]
Combining the three contributions gives $\E_{L,R,R'}[I_v\mathcal W_v]=O(1)$ for every fixed value of $l(v)$.
Averaging over \(l(v)\) gives the same
bound without conditioning. By symmetry, the expected cost of a possible
invocation of \(\textsc{Recolor}(u)\) is also \(O(1)\). Therefore,
lines~4--5 of \cref{alg:insertedge} take \(O(1)\) expected time, completing
the proof.
\end{proof}

\section{Parallel Batch-Dynamic Algorithm}\label{sec:parallel}

In this section, we prove \cref{thm:pram-main}: 

\pramA*

To prove \cref{thm:pram-main}, we introduce a parallel variant of the algorithm of \cref{thm:sequential-main}. Then we prove a series of results that shows that this variant is highly parallel while being work efficient. 

\subsection{Algorithm}\label{sec:parallel_base}
We adopt the definitions, data structures, and initialization procedures as in \cref{sec:sequential}.

After every update, we maintain the data structures $\textsc{LowerEqualUsed}$, $NB_{\geq}$, and $NB_{>}$ in $O(1)$ expected time. For the batch deletion, this is the only work performed.

For a batch insertion, we form a set \(S\) of vertices to be recolored.  For
each inserted edge \((u,v)\), if the two endpoints currently have the same
color, we insert into \(S\) the endpoint that is larger in the lexicographic
order of \((l(\cdot),t(\cdot))\).  In addition, every endpoint of an inserted
edge is independently refreshed with probability \(10^{-4}/\theta(\cdot)\); if
the refresh succeeds, the endpoint is also added to \(S\).  We then call
\(\textsc{RecolorBatch}(S)\).

The main difference from the sequential procedure is that an entire set of
vertices may be recolored simultaneously.  This is handled by the auxiliary
procedure \(\textsc{FixColors}(S)\), whose purpose is to assign new colors to
vertices in \(S\). The recoloring procedure for \(S\) is as follows.  First, every colored vertex in
\(S\) gives up its old color.  These old colors are removed from the relevant
\(\textsc{LowerEqualUsed}\) data structures, so that they no longer constrain
the new samples.  The procedure then samples, for each \(v\in S\), a uniformly
random color from \(\mathcal A_v\), where \(\mathcal A_v\) is defined with
respect to the currently colored vertices and ignores vertices that are
temporarily uncolored.  Thus the sampled color of \(v\) avoids all currently
colored neighbors whose level is at most \(l(v)\).

However, since the vertices in \(S\) sample their colors in parallel, two
adjacent vertices in \(S\) may still sample the same color.  Such a same-colored
edge is resolved directionally.  If the endpoints have different levels, the
higher-level endpoint loses its sampled color.  If the endpoints have the same
level, then both endpoints lose their sampled colors.  After this step, the set
\(S\) is partitioned into two sets \(S_{succ}\) and \(S_{fail}\): vertices in
\(S_{succ}\) keep their sampled colors, while vertices in \(S_{fail}\) remain
temporarily uncolored and must be recolored again.

Once \(\textsc{FixColors}(S)\) finishes, every vertex in \(S_{succ}\) has a new
committed color, and these committed colors are consistent with all
lower-or-equal-level constraints.  We then assign new timestamps to the vertices
in \(S_{succ}\).  Since the colors are sampled in parallel, ties are broken, for
example by vertex index, so that all timestamps remain distinct.

Finally, for each successfully recolored vertex \(v\in S_{succ}\), we propagate
the effect of its new color to higher-level neighbors.  If a higher-level
neighbor now has the same color as \(v\), that neighbor is added to the next
recoloring set.  In addition, each higher-level neighbor is independently added
to the next recoloring set with probability \(10^{-4}/\theta(\cdot)\).  We also
include every vertex in \(S_{fail}\), whose color has not yet been fixed.  The
procedure then recursively recolors this next set.
We present the pseudocodes below.

\begin{algorithm}[H]
\caption{Procedure $\textsc{InsertEdgeBatch}(B)$}\label{alg:insertedgebatch}
\begin{algorithmic}[1]
\small
\State Update $\textsc{LowerEqualUsed}, NB_\geq, NB_>$.
\State $S = \emptyset, R = \emptyset$. \Comment{$R$ is a multiset}
\For{$(u, v) \in B$ in parallel}
\If{$(l(u),t(u))>_{\mathrm{lex}}(l(v),t(v))$} \Comment{Lexicographic comparison of pairs}
\State $\textsc{Swap}(u, v)$
\EndIf
\If{$c(u) = c(v)$}
\State $S = S \cup \{v\}$
\EndIf
\State $R = R \cup \{u,v\}$
\EndFor
\For{$v \in R$ in parallel}
\If {$\textsc{WithProbability}(\frac{10^{-4}}{\theta(v)})$}
\State $S = S \cup \{v\}$
\EndIf
\EndFor
\State $\textsc{RecolorBatch}(S)$
\end{algorithmic}
\end{algorithm}
\begin{algorithm}[H]
\caption{Procedure $\textsc{DeleteEdgeBatch}(B)$}\label{alg:deleteedgebatch}
\begin{algorithmic}[1]
\small
\State Update $\textsc{LowerEqualUsed}, NB_\geq, NB_>$.
\end{algorithmic}
\end{algorithm}

\begin{algorithm}[H]
\caption{Procedure $\textsc{RecolorBatch}(S)$}\label{alg:recolorbatch}
\begin{algorithmic}[1]
\small
\If{$S=\emptyset$}
\State \Return
\EndIf
\State $(S_{succ}, S_{fail}) = \textsc{FixColors}(S)$
\State $S_{next} = S_{fail}$
\For{$v \in S_{succ}$ in parallel}
\For{$w \in NB_{>}(v)$ in parallel}
\If {$c(v) = c(w)$}
\State $S_{next} = S_{next} \cup \{w\}$
\EndIf
\If {$\textsc{WithProbability}(\frac{10^{-4}}{\theta(w)})$}
\State $S_{next} = S_{next} \cup \{w\}$
\EndIf
\EndFor
\EndFor
\State $\textsc{RecolorBatch}(S_{next})$\end{algorithmic}
\end{algorithm}

\begin{algorithm}[H]
\caption{Procedure $\textsc{FixColors}(S)$}\label{alg:fixcolors}
\begin{algorithmic}[1]
\small
\For{$v \in S, c(v) \neq \bot$ in parallel}
\For{$w \in NB_{\geq}(v)$ in parallel}
\State $\textsc{LowerEqualUsed}(w).\textsc{Delete}(c(v))$
\EndFor
\State $c(v) = \bot$
\EndFor
\For{$v \in S$ in parallel}
\State $c(v) = \textsc{LowerEqualUsed}(v).\textsc{SampleNotIn}()$
\EndFor
\State $S_{fail} = \emptyset$
\For{$v \in S$ in parallel}
\For{$w \in NB_{\geq}(v)$ in parallel}
\If{$w \in S$ and $c(v) = c(w)$}
\State $S_{fail}.\textsc{Insert}(w)$
\EndIf
\EndFor
\EndFor
\For{$v \in S_{fail}$ in parallel}
\State $c(v) = \bot$
\EndFor
\For{$v \in S \setminus S_{fail}$ in parallel}
\State Update the timestamp $t(v)$ to the current time, tiebreaking so that every $t(*)$ is distinct.
\For{$w \in NB_{\geq}(v)$ in parallel}
\State $\textsc{LowerEqualUsed}(w).\textsc{Insert}(c(v))$
\EndFor
\EndFor
\Return $(S \setminus S_{fail}, S_{fail})$
\end{algorithmic}
\end{algorithm}

\subsection{Collision Probability}
Throughout this section, we treat the level assignment as fixed and arbitrary. In particular, the arguments in this section do not use any randomness from the level assignment.

As in \cref{obs:hierarchy} of \cref{sec:sequential}, our algorithm satisfies the property that the behavior of a vertex $v$ is not affected by random choices made at vertices of a higher level. 

\begin{observation}\label{obs:hierarchypar}
Fix the graph, the update sequence, and the level assignment. At any given time, the color of a vertex $v$ is a deterministic function of these fixed objects together with the random choices made at vertices of level at most $l(v)$. Here, a random choices made at a vertex $v$ consists of the invocation of $\textsc{LowerEqualUsed}(v).\textsc{SampleNotIn}()$ and $\textsc{WithProbability}(\frac{10^{-4}}{\theta(v)})$. If we know all random choices made over all vertices $w$ with $l(w)\le l(v)$, then the color of $v$ is determined. In particular, the color of $v$ is independent of all random choices made at vertices whose level is strictly larger than $l(v)$.
\end{observation}

\begin{definition}\label{def:goodlevelbatch}
    Fix the graph, update sequence, a vertex $v$ and its level $l(v)$. We call the level assignment $l$ to be \textbf{good} for $v$ if it satisfies $(\Delta+1-d_{\leq}(v))\ge \theta(v)$ throughout the last $2\theta(v)^2$ batch edge insertions that contains an edge incident to $v$. We call the level assignment $l$ to be \textbf{bad} for $v$ if is not good for $v$. This property is determined solely by the input sequence and the level assignment. In particular, it is independent of the random choices made by the recoloring procedure.
\end{definition}

We show that \cref{lem:jump} and \cref{lem:jump2} applies to our parallel algorithm as well.

\begin{lemma}\label{lem:jumpbatch}
Fix the following objects in order, where each object may depend on the preceding ones:
\begin{itemize}
\item the graph, initially without edges;
\item the update sequence;
\item the level assignment;
\item a vertex $v$ such that the level assignment is good regarding $v$;
\item all random choices made at vertices whose level is less than $l(v)$;
\item a color $q^* \in[\Delta+1]$.
\end{itemize}
Then, over the remaining random choices made at vertices whose level is at least $l(v)$, the probability that $c(v)=q^*$ at the end of the update sequence is at most $O(1/\theta(v))$.
\end{lemma}

\begin{lemma}\label{lem:jump2batch}
Fix the following objects in order, where each object may depend on the preceding ones:
\begin{itemize}
\item the graph, initially without edges;
\item the update sequence;
\item the level assignment;
\item a vertex $u$;
\item a vertex $v$ such that the level assignment is good regarding $v$;
\item all random choices made at vertices whose level is less than $l(v)$.
\end{itemize}
Then, over the remaining random choices made at vertices whose level is at least $l(v)$, the probability that $c(u)=c(v)$ and $t(u)<t(v)$ at the end of the update sequence is at most $O(1/\theta(v))$.
\end{lemma}

To prove \cref{lem:jumpbatch} and \cref{lem:jump2batch}, we follow the observation used in \cref{sec:sequential}.

Define the following two types of events that are \emph{relevant} to $v$:
\begin{itemize}
\item \textbf{Type A:} an edge incident to $v$ is inserted;
\item \textbf{Type B:} a neighbor of $v$ whose level is less than $l(v)$ resamples its color in \cref{alg:fixcolors}.
\end{itemize}

For both lemmas, the fixed context determines the full sequence of events relevant to $v$. Type A events are determined directly by the update sequence. Type B events are determined by \cref{obs:hierarchypar}, since they involve only vertices of level less than $l(v)$. Let $R_1, R_2, \ldots, R_k$ denote the resulting sequence of relevant events in the order of occurrence. 

Compared to the sequential algorithm, there are two distinctions to keep in mind for the parallel algorithm.

First, a single invocation of \cref{alg:recolorbatch} may generate multiple
relevant events. We impose an arbitrary order on the relevant events generated within that invocation.  For the analysis, we consider a virtual
sequential process in which the recoloring requests are processed one by one according to this imposed order. In the actual algorithm, there is no sequential order and duplicate requests made during the same
function call are merged. However, this virtual sequential process is still equivalent to the actual implementation: recoloring a vertex once is equivalent to recoloring it several times during the same function call and keeping only the last sampled color.

Second, \cref{alg:fixcolors} does not necessarily assign committed colors to all vertices in \(S\).  Thus, even if a recoloring request is generated at a specific relevant event, the actual color may be assigned only at some later time.  Until then, the vertex is \emph{uncolored}.  Accordingly, in the proof we do not assume that a new color is assigned immediately after each recoloring stage.  Instead, each recoloring will be completed at a later point which is between two consecutive relevant events. If the completed recoloring represents the merger of several recoloring requests, we regard all of those requests as being processed at this same completion time.
As such, vertices in \(S_{fail}\) in \cref{alg:recolorbatch} are not treated as newly inserted into \(S\).  They were already in \(S\); they
simply remain pending because their attempted recoloring failed inside
\cref{alg:fixcolors}.  

For $v$ to be in $S$ in \cref{alg:fixcolors}, one of the following should hold:
\begin{itemize}
\item An edge incident to $v$ is inserted, and by the line 7 of \cref{alg:insertedgebatch}, $v$ is added into $S$ that would be batch-recolored. This event can occur only during a Type A event.

\item An edge incident to $v$ is inserted, and by the line 11 of \cref{alg:insertedgebatch}, $v$ is added into $S$ that would be batch-recolored. This event occurs exactly when a Type A event occurs and the call to $\textsc{WithProbability}$ in line 10 of \cref{alg:insertedgebatch} returns $\textsf{TRUE}$.

\item An edge not incident to $v$ is inserted, and by line 8 of \cref{alg:recolorbatch}, $v$ is added into $S_{next}$ that would be batch-recolored. This event can occur only during a Type B event.

\item An edge not incident to $v$ is inserted, and by line 10 of \cref{alg:recolorbatch}, $v$ is added into $S_{next}$ that would be batch-recolored. This event occurs exactly when a Type B event occurs and the call to $\textsc{WithProbability}$ in line 9 of \cref{alg:recolorbatch} returns $\textsf{TRUE}$.
\end{itemize}

Thus, relevant events account for every possible recoloring request for \(v\).
Moreover, each relevant event can generate at most one direct recoloring request,
and it also generates exactly one probabilistic recoloring opportunity, namely
the request that is made if the corresponding \(\textsc{WithProbability}\) call
returns \(\textsf{TRUE}\). We therefore view each relevant event as having two
stages: first a direct recoloring stage, followed by a probabilistic recoloring stage. 

The following property is obvious in \cref{sec:sequential}, but is less so in the parallel algorithm. Nonetheless we show that this property holds.

\begin{lemma}\label{lem:fixcolor_initialprob}
    Fix a vertex $v$ and a color $q \in [\Delta + 1]$. For any vertex $v$ that is recolored, $\Pr[c(v) = q] \le \frac{1}{\Delta + 1 - d_\leq(v)}$, where the probability is only over the randomization to sample from the set $\mathcal A_v$ in line 6 of \cref{alg:fixcolors}. All other random choices may be fixed arbitrarily.
\end{lemma}

\begin{proof}
Fix all random choices except for the sample of \(v\) from \(\mathcal A_v\) in
line 6 of \cref{alg:fixcolors}.  We condition on each function call that successfully assigns a non-conflicting color to $v$. 

There are two kinds of colors that can create a conflict for \(v\) .
First, \(v\) cannot use a color already held by a currently colored neighbor of level at most \(l(v)\); these colors are excluded from \(\mathcal A_v\).  Second, among the neighbors that are temporarily uncolored and are sampled in parallel with \(v\), a lower-or-equal-level neighbor may sample the same color as \(v\). Since all random choices other than \(v\)'s sample are fixed, these colors are
also fixed.

Altogether, at most \(d_{\leq}(v)\) colors are forbidden by lower-or-equal-level
neighbors of \(v\).  Hence the set of colors from which \(v\)'s sample can be
kept has size at least $\Delta+1-d_{\leq}(v)$.
Conditioned on the sample being kept, the color of \(v\) is uniform over this
set of admissible colors.  Therefore, for any fixed color \(q\), the conditional
probability that \(c(v)=q\) is either \(0\), if \(q\) is not admissible, or at
most $\frac{1}{\Delta+1-d_{\leq}(v)}$.
\end{proof}

To prove \cref{lem:jumpbatch}, we need the following sublemma.

\begin{lemma}\label{lem:jumpbatch_sublemma}
Fix the context of \cref{lem:jumpbatch}.  Let \(R_1,\ldots,R_k\) be the
resulting sequence of relevant events.  We also write \(R_0\) for the initial
coloring stage of \(v\), and \(R_{k+1}\) for the end of the process.  In
addition, fix the following:
\begin{itemize}
    \item an index \(j\in\{0,1,\ldots,k\}\);
    \item a recoloring of \(v\) that completes after \(R_j\) and before \(R_{j+1}\);
    \item all random choices made at vertices of level exactly \(l(v)\) before
    this recoloring of \(v\).  When \(j=0\), this means the initial colors of
    all vertices of level exactly \(l(v)\), except for \(v\) itself.  If several vertices are recolored simultaneously, we respect the timestamp
    tie-breaking order: among such vertices, we also fix the random choices made by vertices whose timestamps are smaller than the timestamp assigned to
    \(v\).
\end{itemize}

Condition on the event that no subsequent probabilistic recoloring of \(v\) is
triggered after this recoloring completes.  Then the probability that
\(c(v)=q^*\) at any point after this recoloring of \(v\) is at most
\[
\frac{2(k-j)/\theta(v)+1}{\theta(v)}.
\]
\end{lemma}

\begin{proof}
We may assume that $k-j\le \theta(v)^2$, since otherwise the desired bound is trivial. Under this assumption, the condition $\Delta+1-d_{\leq}(v)\ge \theta(v)$ holds at every moment from $R_j$ and onward. To see this, first note that edge deletions cannot violate the condition: if the condition holds before deleting an edge, then it also holds afterward, since $d_{\leq}(v)$ can only decrease. Hence the last moment at which the condition could become violated is immediately after a batch edge insertion. By \cref{def:goodlevelbatch}, and the fact that every batch edge insertion accounts for at least one Type A event, this occurs before $R_j$. Thus the condition holds throughout the interval we analyze.

Immediately after $v$ is recolored, the probability that $c(v)=q^*$ is at most $1/\theta(v)$, by \cref{lem:fixcolor_initialprob}.

Suppose that $c(v)\neq q^*$ immediately after this recoloring. Then $c(v)$ can become equal to $q^*$ only if $v$ is recolored again. Since we have conditioned on the event that none of the subsequent probabilistic recoloring of $v$ is triggered, and any prior recoloring requests from $R_1, R_2, \ldots, R_j$ are completed either by this recoloring of $v$ or the older ones, any later recoloring of $v$ must be forced directly by one of the events $R_{j+1},R_{j+2},\ldots,R_k.$

We now identify a necessary condition for each such event to force the next recoloring of $v$.

First consider a Type A event $R_{j+t}$, where $t\ge 1$. Let the inserted edge be $(v,w)$. We claim that there is a color $q_{j+t}$, determined by the current context, such that $R_{j+t}$ can force the next recoloring of $v$ only if $c(v)=q_{j+t}$ immediately before $R_{j+t}$. Consider the three possible comparisons between $l(w)$ and $l(v)$:

\begin{itemize}
\item If $l(w)<l(v)$, then the conflicting color is simply the color of $w$ at the moment of the insertion. This color is determined by the fixed context. If $w$ is uncolored, set $q_{j + t}$ arbitrarily.
\item If $l(w)=l(v)$, then the only possible conflicting color is the color of $w$ immediately before the recoloring of $v$. Indeed, suppose that $v$ is forced to recolor at $R_{j+t}$ even though its color does not match the color that $w$ had immediately before the recoloring of $v$. Then $w$ must have been recolored after the recoloring of $v$. In that case, $t(w)>t(v)$, so the algorithm would recolor $w$, not $v$. Therefore this case cannot force a recoloring of $v$ and we reach a contradiction. Hence, when $l(w)=l(v)$, the relevant conflicting color is already determined before the recoloring of $v$, and is therefore fixed by the current context. If $w$ is uncolored, set $q_{j + t}$ arbitrarily.
\item If $l(w)>l(v)$, then the insertion cannot force $v$ to recolor, since the algorithm recolors toward the higher-level endpoint. Thus we can set $q_{j + t}$ arbitrarily.
\end{itemize}

Now consider a Type B event $R_{j+t}$, again with $t\ge 1$. The same conclusion holds: there is a color $q_{j+t}$, determined by the current context, such that $R_{j+t}$ can force the next recoloring of $v$ only if $c(v)=q_{j+t}$. In this case, $q_{j+t}$ is simply the new color of the lower-level neighbor $w$ of $v$ that has just been recolored, and this color is determined by the fixed context. If $w$ is uncolored, set $q_{j + t}$ arbitrarily.

We prove the lemma by induction on the decreasing order of $j$. Order the colors in $\mathcal A_v\setminus\{q^*\}$ by the first time they appear among
\[
q_{j+1},q_{j+2},\ldots,q_k.
\]
Colors that never appear in this sequence may be placed arbitrarily after all colors that do appear. 

Suppose the color sampled for $v$ is the $i$-th color in this order. We claim that in this case, the probability that $c(v)=q^*$ at some later point is at most
\[
\frac{2(k-j-i)/\theta(v)+1}{\theta(v)}
\]
when $j+i\le k$, and is $0$ otherwise. The induction hypothesis applies uniformly over all valid continuations of the random choices, as long as the corresponding event triggers a direct recoloring of $v$. Hence, we can condition on any randomization that had been chosen after $R_j$. Although we don't know what is the actual recoloring event that is triggered (if there is any), we know that it is no earlier than $R_{j + i}$, and among all valid possibilities the upper bound of $R_{j + i}$ dominates

Let
\[
r=\min(\theta(v)-1,k-j).
\]
The total probability is at most
\[
\begin{aligned}
&\frac{1}{\theta(v)}
+
\sum_{i=1}^{r}
\frac{1}{\theta(v)}
\cdot
\frac{2(k-j-i)/\theta(v)+1}{\theta(v)}
\end{aligned}
\]

As shown in the proof of \cref{lem:jump_sublemma}, this expression is upper bounded by $\frac{2(k-j)/\theta(v)+1}{\theta(v)}$.
\end{proof}

To prove \cref{lem:jump2batch}, we also show the analogue of \cref{lem:jumpbatch_sublemma}.
\begin{lemma}\label{lem:jump2batch_sublemma}
Fix the context of \cref{lem:jump2batch}.  Let \(R_1,\ldots,R_k\) be the
resulting sequence of relevant events.  We also write \(R_0\) for the initial
coloring stage of \(v\), and \(R_{k+1}\) for the end of the process.  In
addition, fix the following:
\begin{itemize}
    \item an index \(j\in\{0,1,\ldots,k\}\);
    \item a recoloring of \(v\) that completes after \(R_j\) and before \(R_{j+1}\);
    \item all random choices made at vertices of level exactly \(l(v)\) before
    this recoloring of \(v\).  When \(j=0\), this means the initial colors of
    all vertices of level exactly \(l(v)\), except for \(v\) itself.  If several vertices are recolored simultaneously, we respect the timestamp
    tie-breaking order: among such vertices, we also fix the random choices made by vertices whose timestamps are smaller than the timestamp assigned to
    \(v\).
\end{itemize}

Condition on the event that no subsequent probabilistic recoloring of \(v\) is
triggered after this recoloring completes.  Then the probability that
$c(u) = c(v)$ and $t(u) < t(v)$ at any point after this recoloring of \(v\) is at most
\[
\frac{2(k-j)/\theta(v)+1}{\theta(v)}.
\]
\end{lemma}

\begin{proof}
The difference between \cref{lem:jumpbatch_sublemma} and the present lemma is that we now bound the event
\[
c(u)=c(v) \quad\text{and}\quad t(u)<t(v)
\]
where $u$ is a vertex at the same level as $v$, instead of bounding the event $c(v)=q^*$ for a fixed color $q^*$.

We may assume that $k-j\le \theta(v)^2$, since otherwise the desired bound is trivial. Under this assumption, the condition $\Delta+1-d_{\leq}(v)\ge \theta(v)$ holds at every moment from $R_j$ and onward. To see this, first note that edge deletions cannot violate the condition: if the condition holds before deleting an edge, then it also holds afterward, since $d_{\leq}(v)$ can only decrease. Hence the last moment at which the condition could become violated is immediately after a batch edge insertion. By \cref{def:goodlevelbatch}, and the fact that every batch edge insertion accounts for at least one Type A event, this occurs before $R_j$. Thus the condition holds throughout the interval we analyze.

Immediately after $v$ is recolored, the probability that $c(v)=c(u)$ is at most $1/\theta(v)$, by \cref{lem:fixcolor_initialprob}. Here $c(u)$ is fixed by the current context. Since we only need an upper bound, we do not need to use the condition $t(u)<t(v)$ if this case happens.

Suppose that $c(v)\neq c(u)$ immediately after this recoloring. Then the event $c(u)=c(v)$ and $t(u)<t(v)$ can occur only after $v$ is recolored again. Indeed, recoloring $u$ may make $c(u)=c(v)$ true, but if this happens without a later recoloring of $v$, then the timestamp of $u$ becomes larger than the timestamp of $v$, violating the condition $t(u)<t(v)$. Since we have conditioned on the event that none of the subsequent probabilistic recoloring of $v$ is triggered, and any prior recoloring requests from $R_1, R_2, \ldots, R_j$ are completed either by this recoloring of $v$ or the older ones, any later recoloring of $v$ must be forced directly by one of the events $R_{j+1},R_{j+2},\ldots,R_k$.

By identifying a necessary condition under which each such event can force the
next recoloring of \(v\), we reduce to exactly the setting considered in the
proof of \cref{lem:jumpbatch_sublemma}.  The same inductive argument then yields the
upper bound $\frac{2(k-j)/\theta(v)+1}{\theta(v)}$.
\end{proof}

We now prove \cref{lem:jumpbatch} and \cref{lem:jump2batch}.

\begin{proof}[Proof of \cref{lem:jumpbatch}]
Let $J$ be the index of the last relevant event that triggers a probabilistic recoloring of $v$. If no relevant event triggers a probabilistic recoloring of $v$, set $J=0$.

We bound
\[
\sum_{j=0}^{k}
\Pr[c(v)=q^* \mid J=j]\cdot \Pr[J=j].
\]
The probabilistic recoloring of $v$ would complete after an event $R_{j'}$, where $j' \geq j$. As the upper bound in \cref{lem:jumpbatch_sublemma} is decreasing over $j$ and works for any fixed context, we can take $\frac{2(k-j)/\theta(v)+1}{\theta(v)}$ as an upper bound for $\Pr[c(v)=q^* \mid J=j]$. This is at most
\[
\frac{2k/\theta(v)+1}{\theta(v)}
\left(1-\frac{10^{-4}}{\theta(v)}\right)^k
+
\sum_{j=1}^{k}
\frac{2(k-j)/\theta(v)+1}{\theta(v)}
\cdot
\frac{10^{-4}}{\theta(v)}
\left(1-\frac{10^{-4}}{\theta(v)}\right)^{k-j}.
\]

As shown in the proof of \cref{lem:jump}, this expression is upper bounded by $O(\frac{1}{\theta(v)})$.
\end{proof}

\begin{proof}[Proof of \cref{lem:jump2batch}]
Let $J$ be the index of the last relevant event that triggers a probabilistic recoloring of $v$. If no relevant event triggers a probabilistic recoloring of $v$, set $J=0$.

We bound
\[
\sum_{j=0}^{k}
\Pr[c(u)=c(v)\land t(u)<t(v) \mid J=j]\cdot \Pr[J=j].
\]
The probabilistic recoloring of $v$ would complete after an event $R_{j'}$, where $j' \geq j$. As the upper bound in \cref{lem:jump2batch_sublemma} is decreasing over $j$ and works for any fixed context, we can take $\frac{2(k-j)/\theta(v)+1}{\theta(v)}$ as an upper bound for $\Pr[c(u)=c(v)\land t(u)<t(v) \mid J=j]$. This is at most
\[
\frac{2k/\theta(v)+1}{\theta(v)}
\left(1-\frac{10^{-4}}{\theta(v)}\right)^k
+
\sum_{j=1}^{k}
\frac{2(k-j)/\theta(v)+1}{\theta(v)}
\cdot
\frac{10^{-4}}{\theta(v)}
\left(1-\frac{10^{-4}}{\theta(v)}\right)^{k-j}.
\]

As shown in the proof of \cref{lem:jump}, this expression is upper bounded by $O(\frac{1}{\theta(v)})$.
\end{proof}

\begin{lemma}\label{lem:EVcollideCbatch}
For any edge \((u,v)\) in a batch insertion, and for any level assignment $l$ that is good for \(v\), the
probability that
\[
(l(u),t(u))<_{\mathrm{lex}}(l(v),t(v))
\qquad\text{and}\qquad
c(u)=c(v)
\]
immediately before the insertion is \(O(1/\theta(v))\).
\end{lemma}

\begin{proof}
Fix the graph and the prefix of the update sequence ending immediately
before the insertion of batch that contains an edge \((u,v)\). Consider the execution of the data
structure on this fixed prefix, conditioned on the level assignment being $l$. Fix all
random choices made at vertices whose level is less than \(l(v)\).

If \(l(u)<l(v)\), by \cref{obs:hierarchypar}, the color \(c(u)\)
is fixed under the current conditioning; denote it by \(q^*\). Since \(l\)
is good for \(v\), \cref{lem:jumpbatch} implies that, over the remaining
randomness, $\Pr[c(v)=q^*]=O(1/\theta(v))$. 
Thus, in this case, the probability that
\((l(u),t(u))<_{\mathrm{lex}}(l(v),t(v))\) and \(c(u)=c(v)\) is
\(O(1/\theta(v))\).

If \(l(u)=l(v)\), since \(l\) is good for \(v\),
\cref{lem:jump2batch} gives $\Pr[t(u)<t(v)\land c(u)=c(v)]=O(1/\theta(v))$. 

Therefore, for every fixed outcome of the random choices made below level \(l(v)\),
the probability in the statement is \(O(1/\theta(v))\). Averaging over the random choices made below level \(l(v)\)
preserves this bound.
\end{proof}

\subsection{Expected Work For Each Recoloring}
We next bound the expected running time of \(\textsc{RecolorBatch}\), including
all recursive calls that it triggers. We begin by showing that in each call to \(\textsc{FixColors}(S)\), every vertex has a
constant probability of keeping its sampled color.

\begin{lemma}\label{lem:constantprob}
Fix a set of vertices \(S\subseteq V\) and a vertex \(v\in S\).  After one call
to \(\textsc{FixColors}(S)\), the probability that \(v\in S_{succ}\) is at
least $10^{-3}$.  The probability is taken only over the colors sampled by the
vertices in $\{v\}\cup (NB_{\leq}(v)\cap S)$
during this call to \(\textsc{FixColors}(S)\); all other random choices are
fixed arbitrarily.
\end{lemma}

\begin{proof}
Let $W = S\cap NB_{\leq}(v)$. Fix all random choices except for the colors sampled by the vertices in $W \cup \{v\}$.
For a color \(q\in \mathcal A_v\), condition on the event that \(v\) samples
\(q\).  Then \(v\)'s sampled color is rejected only if some vertex
\(w\in W\) also samples \(q\).  Therefore, the probability that \(q\) avoids all
colors sampled by vertices in \(W\) is
\[
\prod_{\substack{
    w\in W\\
    q\in\mathcal A_w
}}
\left(1-\frac{1}{|\mathcal A_w|}\right).
\]
Using the inequality
\[
1-\frac{1}{x}\ge \exp\left(-\frac{2}{x}\right)
\qquad\text{for all }x\ge 2,
\]
and the fact that \(|\mathcal A_w|\ge 2\) in this setting, since \(v\) is
temporarily uncolored, this product is at least
\[
\exp\left(
-2\sum_{\substack{
    w\in W\\
    q\in\mathcal A_w
}}
\frac{1}{|\mathcal A_w|}
\right).
\]

Define
\[
f(q)=
\sum_{\substack{
    w\in W\\
    q\in\mathcal A_w
}}
\frac{1}{|\mathcal A_w|}.
\]
We call a color \(q\in\mathcal A_v\) \emph{light} if \(f(q)\le 2\).  By the
bound above, if \(v\) samples a light color \(q\), then the sampled color is kept
with probability at least $\exp(-4)$. We show that a constant fraction of the colors in \(\mathcal A_v\)
are light.

First, we bound the average value of \(f(q)\) over \(q\in\mathcal A_v\).  We
have
\[
\sum_{q\in\mathcal A_v} f(q)
=
\sum_{q\in\mathcal A_v}
\sum_{\substack{
    w\in W\\
    q\in\mathcal A_w
}}
\frac{1}{|\mathcal A_w|}
=
\sum_{w\in W}
\frac{|\mathcal A_v\cap\mathcal A_w|}{|\mathcal A_w|}
\le
|W|.
\]
On the other hand,
\[
|\mathcal A_v|
\ge
\bigl(\Delta+1-|NB_{\leq}(v)\setminus S|\bigr)
\ge
|NB_{\leq}(v)|-|NB_{\leq}(v)\setminus S|
=
|S\cap NB_{\leq}(v)|
=
|W|.
\]
Combining the two inequalities gives
\[
\sum_{q\in\mathcal A_v} f(q)\le |\mathcal A_v|.
\]
Thus the average value of \(f(q)\), over a uniformly random
\(q\in\mathcal A_v\), is at most \(1\).  By Markov's inequality, at most half of
the colors \(q\in\mathcal A_v\) satisfy \(f(q)>2\).  Hence at least half of
the colors in \(\mathcal A_v\) are light.

Therefore, with probability at least \(1/2\), vertex \(v\) samples a light color;
conditioned on this, the sampled color is kept with probability at least
\(\exp(-4)\).  Consequently,
\[
\Pr[v\in S_{succ}]
\ge
\frac{\exp(-4)}{2}
\approx 0.00916
\ge 10^{-3}.
\]
\end{proof}

Next, we introduce the parallel analogue of \cref{lem:EVrecolor}. We first specify how we charge the work to initial recoloring requests when requests at the same vertex merge.

\begin{definition}\label{def:requestownership}
We call the initial request to which we charge a pending recoloring request its \textbf{owner}. Each initial request owns itself. When requests at a vertex merge, we retain one request and its owner according to a fixed rule before the next coloring attempt, and discard the others. Failed attempts retain the pending request and its owner. Requests generated by a successful recoloring inherit the owner of that recoloring. We charge the local work of each attempt to its owner. We use these owners only for the analysis; the parallel algorithm need not store them.
\end{definition}

\begin{definition}\label{def:typicallevelbatch}
We call a level assignment \textbf{typical} for $v$ in the parallel algorithm if it satisfies the conditions of \cref{def:typicallevel}, where we evaluate the graph immediately after inserting the entire current batch and before processing its recolorings. Otherwise, we call the assignment \textbf{atypical} for $v$.
\end{definition}

Fix a vertex $v$ and its level $l(v)$. Let $L$ denote the level assignment on $V\setminus\{v\}$, let $R$ denote the random choices made before the current batch update, and let $R'$ denote the fresh random choices used to process that batch, including its initial refresh choices and all subsequent recoloring attempts. For a potential initial recoloring request at $v$, let $\mathcal W_v$ be the work charged to it under \cref{def:requestownership} when we include it among the initial requests. This includes the work of subsequent requests that inherit its ownership; its contribution is zero if it is discarded in the initial merge. Charging each attempt to its owner accounts for all recoloring work. We bound the work of each owner by bounding all requests it generates, including those subsequently discarded by merging.

\begin{lemma}\label{lem:EVrecolor_par}
For each vertex $v$, in level $l(v)$, the expected work of $\textsc{RecolorBatch}(S)$ incurred by a potential initial recoloring request at $v$ is
\[
\E_{L,R,R'}[\mathcal W_v]=O(\Delta\cdot2^{-l(v)})=O(\theta(v)).
\]
More precisely, let $\mathcal T_v$ be the event that the assignment is typical for $v$, as in \cref{def:typicallevelbatch}. Then:
\begin{enumerate}
    \item For every typical assignment $L=\ell$ and every fixed past randomness $R=r$,
    \[
    \E_{R'}[\mathcal W_v\mid L=\ell,R=r]=O(\theta(v)).
    \]
    We obtain this bound uniformly over the initial state and the other initial recoloring requests in the batch.
    \item The contribution from atypical assignments satisfies
    \[
    \E_{L,R,R'}[\mathbf 1[\lnot\mathcal T_v]\mathcal W_v]=O(1).
    \]
\end{enumerate}
Moreover, the expected-work bound remains valid after conditioning on the event that the assignment is bad for $v$, as defined in \cref{def:goodlevelbatch}.
\end{lemma}

\begin{proof}
We follow the proof of \Cref{lem:EVrecolor}, accounting also for failed coloring attempts. Fix a typical assignment $L=\ell$, any past randomness $R=r$, and any initial collection of recoloring requests. We take expectations over the fresh randomness $R'$ used by the batch recoloring. We prove the bound uniformly over this initial state and over the history preceding each call to $\textsc{FixColors}$.
In the proof of \Cref{lem:EVrecolor}, when a node $u$ is currently uncolored and chooses a color, we argued that, the set of out-neighbors with whom the new color of $u$ collides and are forced to recolor has size at most $1$, in expectation. This was central to the analysis of the recursive cost function $Q(i)$. The reason for that expected upper bound was simple, in the sequential setting: if the token-holder node $u$ has $t$ out-neighbors, it chooses its random color from a space of at least $t+1$ colors, and hence the number of out-neighbors who are forced to be recolored is $\frac{t}{t+1}<1$, in expectation. In the parallel setting, when a token-holder node $u$ samples a color, simultaneously, other token-holder nodes that are in-neighbors of $u$ or same-level neighbors of it are also trying to sample a color. If any of those chose the same color, the sampled color of $u$ is not successful, and node $u$ tries again to find another color. The color that $u$ eventually receives, in the first time that it is successful, does not come from the same simple uniform distribution over $t+1$ colors, as in the sequential setting. Despite that, we argue that the expected number of out-neighbors that are forced to recolor (with the first successful color of $u$) is at most $1$. We do this by considering any setting with any number of competing in-neighbors and same-level neighbors, who are trying colors from their own lists.

Let $IN(u) = NB_\leq(u) \cap S$ be the set of in-neighbors or same-level neighbors which are uncolored. Consider a node $u$ where $d = |NB_>(u)|$ is the number of out-neighbors and $d' = |IN(u)|$. Then $|\mathcal A_u| \geq d + d' + 1$. Suppose that each $w \in IN(u)$ picks its color uniformly at random from a list $\mathcal A_w$ with $|\mathcal A_w|\ge 2$, while the out-neighbors are colored adversarially, oblivious to the randomness of nodes in $\{u\} \cup IN(u)$.

\begin{claim}\label{clm:collisionExpectation}
Conditioned on the event that node $u$ samples a color different from the colors sampled by $IN(u)$, the expected number of out-neighbors whose colors conflict with the color of $u$ is less than $1$.
\end{claim}

\begin{proof}
Fix the colors of the out-neighbors. For each color $c\in \mathcal A_u$, let $o_c$ denote the number of out-neighbors colored with $c$. Then $\sum_{c\in \mathcal A_u} o_c \le d$.
Let $o$ denote the number of out-neighbors conflicting with the color sampled by $u$, and let $A=\{\text{\(u\) samples a color that no vertex in \(IN(u)\) samples}\}.$
For each color $c$, let $g_c$ denote the probability that none of the vertex in $IN(u)$ sample color $c$. Then, we have
$
\mathbb{E}[o\mid A]
 = \sum_{c\in \mathcal A_u} o_c \cdot \frac{g_c}{Z},
$ where $Z=\sum_{c\in \mathcal A_u} g_c.$ Hence,
\[
\mathbb{E}[o\mid A]
=
\frac{\sum_{c\in \mathcal A_u} o_c g_c}
{\sum_{c\in \mathcal A_u} g_c}
\le
\frac{\sum_{c\in \mathcal A_u} o_c}
{\sum_{c\in \mathcal A_u} g_c}
\le
\frac{d}
{\sum_{c\in \mathcal A_u} g_c}.
\]
We now analyze the denominator. For each color $c$,
\[
g_c
=
\prod_{\substack{w\in \mathrm{IN}(u)\\ c\in \mathcal A_w}}
\left(1-\frac{1}{|\mathcal A_w|}\right)
\ge
1-
\sum_{\substack{w\in \mathrm{IN}(u)\\ c\in \mathcal A_w}}
\frac{1}{|\mathcal A_w|},
\]
where the inequality follows from the union bound.
Therefore,
\[
\sum_{c\in \mathcal A_u} g_c
\ge
\sum_{c\in \mathcal A_u}
\left(
1-
\sum_{\substack{w\in \mathrm{IN}(u)\\ c\in \mathcal A_w}}
\frac{1}{|\mathcal A_w|}
\right)
\ge
(d+d'+1)-d'
=
d+1.
\]
Consequently,
\[
\mathbb{E}[o\mid A]
\le
\frac{d}{d+1}
<
1.
\]
\end{proof}
We apply \cref{clm:collisionExpectation} after fixing the history preceding each coloring attempt, including the choices from $R'$ already used. Conditional on success, the expected number of higher-level neighbors triggered by a collision is at most one, with the expectation over the fresh choices from $R'$ used in that attempt. Only higher-level neighbors outside the current set $S$ can generate a new collision request; a higher-level neighbor in $S$ that samples the same color is already retained in $S_{fail}$. Averaging over the possible histories preceding the eventual successful attempt gives the same bound for that attempt. Requests discarded by merging generate no further work of their own.

By \cref{lem:constantprob}, each attempt succeeds with probability at least $10^{-3}$, conditional on its preceding history. Thus, a pending request incurs at most $10^3$ attempts in expectation before success or removal by merging. Failed attempts incur local work but generate no new requests. We therefore increase the constant in the local cost term and the high-level bound $K$ in the proof of \cref{lem:EVrecolor}. The expected number of collision children remains at most one, and the refresh probabilities are unchanged. We apply the recurrence $Q(i)$ from that proof with these larger cost constants, uniformly over the history preceding each request. All expectations are over the unused choices from $R'$. This gives
\[
\E_{R'}[\mathcal W_v\mid L=\ell,R=r]=O(\theta(v))
\]
for every typical assignment and every fixed past randomness. Since the bound holds for every initial collection of requests, we can also average over the initial refresh choices in $R'$ that determine that collection.

Next, we account for retries on atypical assignments. For a fixed level assignment, we bound the number of requests generated by an owner by allowing every successful recoloring to trigger all higher-level neighbors through both propagation mechanisms. Every chain of generated requests moves to strictly higher levels; failed attempts remain part of the same pending request. The same coarse bounds on the number of requests and their neighborhood scans as in the proof of \cref{lem:EVrecolor} therefore apply. Each request incurs only a constant expected number of attempts over the unused choices from $R'$, uniformly over its preceding history. By linearity of expectation, these coarse bounds give corresponding bounds on expected work. Combining them with \cref{lem:climbbound_claim1} and the per-level concentration estimates in the proof of \cref{lem:EVrecolor}, we obtain
\[
\E_{L,R,R'}[\mathbf 1[\lnot\mathcal T_v]\mathcal W_v]=O(1).
\]
Together, these estimates prove $\E_{L,R,R'}[\mathcal W_v]=O(\theta(v))$. For the bound conditioned on a bad assignment, we use \cref{lem:climbbound_claim3} with the neighborhoods specified by \cref{def:goodlevelbatch}. The conditional per-level estimates in the proof of \cref{lem:EVrecolor} also apply, since the bad-level event remains a decreasing function of membership in $M_v$. The same bounds on requests and expected attempts then prove the conditional expected-work bound.
\par
\end{proof}

\subsection{Efficiency}
\begin{lemma}\label{lem:loground}
    For each batch update, the number of recursive calls to $\textsc{RecolorBatch}$ is at most \roundexact{} with high probability.
\end{lemma}
\begin{proof}
We prove that, with high probability, there is no uncolored node after \roundexact{} calls to $\textsc{RecolorBatch}$. Let us call a level $\ell$ \textit{low} if $\Delta 2^{-\ell} \leq 100\log^2 n$, and \textit{high} otherwise. We first show that, within  $O(\log n)$ calls to $\textsc{RecolorBatch}$, there is no uncolored node left in the high layers, with high probability. Then we show that for any recoloring starting in a low layer, with high probability, it takes at most \roundexact{} additional calls to $\textsc{RecolorBatch}$ until all of the recolorings originating from it have terminated.

We start with the analysis of the high layers. Consider round $i$ to be the configuration after $i$ calls to $\textsc{RecolorBatch}$. Define $U_i$ to be the set of high-layer nodes that are uncolored in round $i$. We define a potential function \[\Phi_i=\sum_{v \in U_i} (1.5)^{\log \Delta - l(v)}.\] Observe that $\Phi_0 \leq n (1.5)^{\log \Delta} \leq n^2$. We prove that $\E[\Phi_{i+1}] \leq (1-10^{-5}) \Phi_i$, where the expectation is only over the randomness in the round $i+1$ call to $\textsc{RecolorBatch}$ and, in particular, it holds for any set $U_i$ at the end of round $i$. As such, we conclude that for a sufficiently large constant $C$, we have $\E[\Phi_{C\log n}] \leq n^{2} (1-10^{-6})^{C\log n} \leq 1/n^{2}$. Hence, by Markov's inequality, after $C\log n$ iterations, with probability at least $1-1/n^{2}$, we have $U_{C\log n} <1$, which means $U_{C\log n}=0$ and thus that no high-layer node remains uncolored. 

Consider round $i+1$, with an arbitrary set $U_i$ of uncolored nodes at its start. Let us focus on an arbitrary node $v\in U_i$. Node $v$ contributes $(1.5)^{\log \Delta - l(v)}$ to the potential $\Phi_i$. This node $v$ may remain uncolored at the end of the $(i+1)^{th}$ call to $\textsc{RecolorBatch}$, in which case its contribution to the potential remains unchanged, or it may receive a new color, in which case it possibly triggers some of its out-neighbors (i.e., nodes $w$ for $l(w)>l(v)$) to be uncolored, thus joining $U_{i+1}$. To analyze the change in the potential, let us have a quick reminder about how the set of out-neighbors that become uncolored is determined. Node $v$, once it gets colored, triggers some of its out-neighbors to be uncolored, in one of two ways: (1) Node $v$ received color $q$ from $\textsc{RecolorBatch}$ as its new successful color, and now each out-neighbor $w$ of $v$ colored with color $q$ gets uncolored and has to recolor itself. (2) Upon the reocloring of $v$, each out-neighbor $w$ of $v$ decides to recolor itself randomly, which happens with probability $\frac{10^{-4}}{\theta(w)}$, where $\theta(w)=\left\lceil \frac{\Delta\cdot 2^{-l(w)}}{100}\right\rceil$.

The probability that $v$ remains uncolored after the $(i+1)^{th}$ call to $\textsc{RecolorBatch}$ is at most $1-10^{-3}$, by \Cref{lem:constantprob}. In the complementary case, where $v$ got colored in that call, we need to understand the set $T(v)$ of out-neighbors that $v$ uncolors, in one of the two ways above, and places in $U_{i+1}$. By \Cref{clm:collisionExpectation}, the expected number of out-neighbors placed in $T(v)$ through the first mechanism (i.e., the color of $v$ colliding with them and thus uncoloring them) is at most $1$. Each such out-neighbor is in a layer $\ell\geq l(v)+1$. Hence, the contribution from this part can be bounded by at most $(1.5)^{\log \Delta - l(v)-1}$. To analyze the second mechanism of spreading, we need to look more closely into the number of neighbors of $v$ in each layer.

Let $\eps=10^{-2}$. Observe that, with high probability, each node $v$ in level $l(v)$ has at most $(1+\eps)\Delta2^{-\ell} + O(\log n)$ out-neighbors in level $\ell$, for each $\ell \geq l(v)+1$. So, for each high level $\ell$, the number of out-neighbors can be bounded as at most $(1+2\eps)\Delta2^{-\ell}$. 

Therefore, we can bound the expected number of out-neighbors of $v$ in each high layer $\ell>l(v)$ that become uncolored when $v$ got colored as at most $(1+2\eps)\Delta2^{-\ell} \cdot \frac{10^{-2}}{\Delta 2^{-\ell}} \leq 0.02$. Hence, in expectation, the contribution of $v$, if it remains uncolored, or the out-neighbors that it uncolors, in case $v$ is colored, can be upper bounded as at most

\begin{align*}& &&(1-10^{-3}) \cdot \Big((1.5)^{\log \Delta - l(v)}\Big) + 10^{-3} \cdot \Big((1.5)^{\log \Delta - l(v)-1} + \sum_{\ell\geq l(v)+1} \big(0.02 \cdot (1.5)^{\log \Delta - \ell}\big)\Big) \\ 
&\leq &&(1-10^{-3}) \cdot \Big((1.5)^{\log \Delta - l(v)}\Big) + 10^{-3} \cdot \Big( (1.5)^{\log \Delta - l(v)} \cdot (\frac{2}{3} + 0.04)\Big) \\
&\leq &&(1-10^{-5}) \cdot (1.5)^{\log \Delta - l(v)}
\end{align*}
The same can be applied to any other node $u\in U_i$, implying that its potential contribution $(1.5)^{\log \Delta-l(u)}$ gets replaced with at most $(1-10^{-5}) \cdot (1.5)^{\log \Delta - l(v)}$ after the $(i+1)^{th}$ call, in expectation. Hence, by the linearity of expectation, we have $\E[\Phi_{i+1}] \leq (1-10^{-5}) \Phi_i$, where the expectation is over the randomness of the $(i+1)^{th}$ call and holds for an arbitrary set $U_i$ of the uncolored nodes right before the $(i+1)^{th}$ call. Therefore, as argued before, after $O(\log n)$ calls, with high probability, there is no uncolored node in any high layer. 

Suppose that there is no uncolored node remaining in high layers. Next, we analyze low layers and prove that after \roundexact{} additional calls to $\textsc{RecolorBatch}$, with high probability, there is no uncolored node in low layers either. Let $\ell^*$ be such that $\Delta2^{-\ell^*}\in (50\log^2n, 100\log^2 n]$. Notice that the low layers are $[\ell^*, \log \Delta]$. From \Cref{lem:constantprob}, we conclude that, with high probability, after $O(\log n)$ additional calls to $\textsc{RecolorBatch}$, there is no uncolored node in layer $\ell^*$. After that, after $O(\log n)$ additional calls to $\textsc{RecolorBatch}$, there is no uncolored node in layers $[\ell^*, \ell^*+1]$, with high probability. And this continues progressing similarly, where after $O(\log n \cdot i)$ calls, we know that there is no uncolored node in layers $[\ell^*, \ell^*+i]$. Since $\log \Delta- \ell^*=O(\log\log n)$, meaning that there are only $O(\log\log n)$ low layers, we conclude that after $O(\log n \cdot \log\log n)$ calls, with high probability, there is no uncolored node in any low layer either, and thus all nodes have been colored.
\end{proof}

\begin{lemma}\label{lem:parallel-base-thm}
For every fixed graph and update sequence, each update takes \(O(1)\)
expected work per edge in the batch, where the expectation is over all random choices made by the algorithm.
\end{lemma}

\begin{proof}
The expected running time of lines~1--11 of \cref{alg:insertedgebatch}, as well
as the entire procedure \cref{alg:deleteedgebatch}, is \(O(1)\) per edge in the
batch.  Thus it remains to bound the expected work of
\(\textsc{RecolorBatch}\), which is invoked in line~12 of
\cref{alg:insertedgebatch}.  We charge this work to each inserted edges and show
that each edge contributes \(O(1)\) expected work. The desired bound then follows by linearity of expectation. 

An inserted edge can contribute to the work of \(\textsc{RecolorBatch}\) in two
ways.  First, if the two endpoints have the same color, line~7 inserts into
\(S\) the endpoint with larger \((l(\cdot),t(\cdot))\).  Second, line~11 may
insert either endpoint into \(S\) through the probabilistic recoloring step.

We first analyze the contribution from the probabilistic recoloring step. Fix an endpoint $v$ and its level $l(v)$, so that $\theta(v)$ is fixed. We use $L$, $R$, and $R'$ for the level assignment, the past randomness, and the fresh randomness, respectively, as in \cref{lem:EVrecolor_par}. For a typical assignment, we condition on $L$, $R$, and the choices from $R'$ preceding the refresh choice. The refresh choice triggers a recoloring with probability $O(1/\theta(v))$. Whenever it triggers, the uniform bound in \cref{lem:EVrecolor_par} bounds its expected subsequent work, over the unused choices from $R'$, by $O(\theta(v))$. Thus, for every typical $\ell$ and every $r$,
\[
\E_{R'}[\text{work attributed to this refresh}\mid L=\ell,R=r]=O(1).
\]
For atypical assignments, we use the exceptional-case work bound from \cref{lem:EVrecolor_par}, which holds uniformly over the initial refresh choices and contributes $O(1)$ to the expectation over $L,R,R'$. Averaging over $l(v)$ and adding the contributions of both endpoints gives $O(1)$ expected work per inserted edge.
\par

It remains to bound the expected work caused by the same-color conflict in
line~7 of \cref{alg:insertedgebatch}. Consider a fixed insertion of an unordered edge \(\{u,v\}\), where \(u\) and
\(v\) denote the two endpoints before the possible reordering in lines~4--5 of
\cref{alg:insertedgebatch}.  Since this reordering depends on the random pairs
\((l(u),t(u))\) and \((l(v),t(v))\), the endpoint that appears second after the
reordering is itself random.  As in the sequential proof, we bound separately
the expected cost of a possible insertion of \(v\) into \(S\); the same argument
applies symmetrically to \(u\).

Fix a possible value \(l'\) of \(l(v)\), and condition on the event
\(l(v)=l'\).  All probabilities and expectations below are taken in this
conditional probability space.  In particular, \(\theta(v)\) is fixed
throughout the remainder of the argument.

Let \(\mathcal P_v\) be the event that, when the inserted edge \(\{u,v\}\) is
processed in \cref{alg:insertedgebatch},
\[
(l(u),t(u))<_{\mathrm{lex}}(l(v),t(v))
\qquad\text{and}\qquad
c(u)=c(v).
\]
Let \(I_v=\mathbf 1[\mathcal P_v]\).  The event \(\mathcal P_v\) is exactly the
condition under which \(v\) is inserted into \(S\) in line~7 of
\cref{alg:insertedgebatch} because of the same-color conflict on the edge
\(\{u,v\}\).

Recall that $\mathcal W_v$ denotes the work attributed to this potential initial recoloring request at $v$, including all subsequent recolorings attributed to it under the ownership convention in \cref{def:requestownership}. Up to an $O(1)$ additive term, we bound the contribution of this request by $I_v\mathcal W_v$. It therefore suffices to prove
\[
\E_{L,R,R'}[I_v\mathcal W_v]=O(1).
\]
\par

Let $\mathcal G_v$ and $\mathcal T_v$ denote the events that the assignment is good and typical for $v$, as in \cref{def:goodlevelbatch,def:typicallevelbatch}. Both events depend only on $L$. We use the same three contributions as in the proof of \cref{thm:sequential-main}: assignments that are good and typical, assignments that are atypical, and assignments that are typical but not good.

First, fix an assignment $L=\ell$ that is both good and typical. Once we fix the past randomness $R$, we determine $I_v$. Using the uniform bound over the fresh randomness $R'$ in \cref{lem:EVrecolor_par}, we obtain
\[
\begin{aligned}
\E_{R,R'}[I_v\mathcal W_v\mid L=\ell]
&=\E_R\!\left[I_v\,\E_{R'}[\mathcal W_v\mid L=\ell,R]\,\middle|\,L=\ell\right]\\
&\le O(\theta(v))\Pr_R[\mathcal P_v\mid L=\ell]\\
&=O(1),
\end{aligned}
\]
where the last equality follows from \cref{lem:EVcollideCbatch}. Averaging over assignments that are both good and typical bounds their contribution by $O(1)$.

For atypical assignments, we use $0\leq I_v\leq1$ and \cref{lem:EVrecolor_par} to obtain
\[
\begin{aligned}
\E_{L,R,R'}[\mathbf 1[\lnot\mathcal T_v]I_v\mathcal W_v]
&\leq\E_{L,R,R'}[\mathbf 1[\lnot\mathcal T_v]\mathcal W_v]\\
&=O(1).
\end{aligned}
\]

Finally, we apply Chernoff's bound and take a union bound over the last $2\theta(v)^2$ relevant batch insertions to obtain
\[
\Pr_L[\lnot\mathcal G_v]\leq2\theta(v)^2\exp(-\Omega(\theta(v))).
\]
The uniform expected-work bound in \cref{lem:EVrecolor_par} applies to every typical assignment, whether or not it is good. Therefore,
\[
\begin{aligned}
\E_{L,R,R'}[\mathbf 1[\lnot\mathcal G_v\cap\mathcal T_v]I_v\mathcal W_v]
&\leq O(\theta(v))\Pr_L[\lnot\mathcal G_v]\\
&\leq O\!\left(\theta(v)^3\exp(-\Omega(\theta(v)))\right)\\
&=O(1).
\end{aligned}
\]
Combining the three contributions gives $\E_{L,R,R'}[I_v\mathcal W_v]=O(1)$ for every fixed value of $l(v)$.
Averaging over \(l(v)\) gives the same bound without conditioning.  By symmetry, the same
bound holds for the possible insertion of \(u\) into \(S\).  Therefore the
same-color conflict in line~7 contributes \(O(1)\) expected work per inserted
edge. 
\end{proof}

\subsection{Proof of \cref{thm:pram-main}}\label{sec:parallel_pram}
In this section, we prove \cref{thm:pram-main} by showing that the algorithm of \cref{sec:parallel_base} admits an efficient \textsf{PRAM} implementation.
\pramA*

\begin{proof}[Proof of \cref{thm:pram-main}]
The expected work bound follows from \cref{lem:parallel-base-thm}, so it remains
to prove the \depthexact{} worst-case depth bound.  The procedures
\cref{alg:insertedgebatch}, \cref{alg:deleteedgebatch},
\cref{alg:fixcolors}, and \cref{alg:recolorbatch} (without the recursion) can each be implemented in \(O(\log n)\) worst-case depth
with high probability.  The \(O(\log n)\) factor comes from the parallel
for-loops, batch updates to hash set implemented using~\cite{gil1991towards}, and the batch updates to \(\textsc{LowerEqualUsed}\)
implemented using \cref{lem:inversesample}.

By \cref{lem:loground}, a call to \cref{alg:recolorbatch} triggered by the
initial batch completes after at most \roundexact{} recursive calls with high
probability.  As each $\textsc{RecolorBatch}$ call can be processed in \(O(\log n)\) depth with high probability, the total depth of the entire recoloring process is \depthexact{} with high probability.
\end{proof}

\section{Distributed Algorithm}\label{sec:parallel_congest}
In this section, we prove \cref{thm:congest-main}.
\congestA*

The algorithm of \cref{sec:parallel} is distributed in nature: each vertex
maintains its own data structures \(\textsc{LowerEqualUsed}\), \(NB_{\geq}\), and \(NB_{>}\) locally, and each edge operation is implemented
by communication between the two endpoints of the edge.  Therefore, the algorithm has a direct (but possibly flawed) implementation in the \textsf{CONGEST} model. 

The main challenge is that the graph may change while a vertex is still waiting
for messages from neighbors defined in a previous round.  In particular, if an
edge is deleted while communication along that edge is pending, the corresponding
communication may fail, and immediately after that the edge could not be used for communication.  Such failures may occur during data-structure
maintenance, edge-insertion processing, or recoloring.  
The simplicity of our
algorithm makes this mostly technical: it turns out that each such failure does not require further action except a purely local operation that only updates local
data structures.

We first consider the simpler setting with no edge deletions.  Let
$T = $ \roundexacttheta{} be chosen sufficiently large so that
\cref{lem:loground} holds with high probability.  We group every consecutive
\(100(T+1)\) rounds into an \emph{epoch}, and record, for each inserted edge,
the epoch in which it was inserted.  If an edge is inserted during epoch \(i\),
we immediately place a token at each endpoint, thereby uncoloring both
endpoints.  This happens unconditionally and is used only to present a valid
partial coloring to the end user.  Internally, the recoloring computation
ignores these tokens, as well as all edges inserted during the current epoch.
At the beginning of epoch \(i+1\), we start acknowledging and processing the
edges inserted during epoch \(i\).

During the first \(100\) rounds of epoch \(i+1\), we simulate
\(\textsc{InsertEdgeBatch}\) from \cref{alg:insertedgebatch}.  For each edge
inserted during epoch \(i\), we create the real tokens corresponding to the
same-color conflict in line~7 and to the probabilistic recoloring in line~11.
These real tokens are a subset of the unconditional tokens created during epoch
\(i\), so from the end user's perspective some of the unconditional tokens
simply disappear.

The next \(100T\) rounds are divided into \(T\) groups of \(100\) rounds.  In
each group, we simulate one call of \(\textsc{FixColors}\)
(\cref{alg:fixcolors}) and \(\textsc{RecolorBatch}\)
(\cref{alg:recolorbatch}).  During this simulation, each token either
disappears, remains at the same vertex, or creates a new token at an adjacent
vertex with the same owner.  These cases correspond respectively to a successful
coloring with no further propagation, a failed coloring attempt, and a
successful coloring that triggers recoloring of a neighbor through lines~8
and~10 of \cref{alg:recolorbatch}.

Throughout these procedures, we ignore edges inserted during epoch \(i+1\).  This
is possible because each edge stores the epoch in which it was inserted.  By the
choice of \(T\) and \cref{lem:loground}, at the end of epoch $i+1$, there would be no tokens from the edges inserted during epoch \(i\), with high
probability.  Thus, in the no-deletion setting, this epoch-based implementation
simulates the parallel algorithm of \cref{sec:parallel}.

We next consider edge deletions.  Deleting an edge cannot invalidate a valid
partial coloring.  Moreover, the procedure \(\textsc{DeleteEdgeBatch}\) is local
and requires no coordination.  The only complication is that, if an edge is
deleted during a round, any communication that was pending along that edge may
be interrupted.
The procedure \(\textsc{InsertEdgeBatch}\) handles this by ignoring all communication failures and checking whether each inserted edge is
still present before invoking \textsc{RecolorBatch}.  If an
edge has already been deleted, we discard it and remove the tokens owned by it.

In \(\textsc{FixColors}\), edge deletions can affect lines~3, 11, and~17 of
\cref{alg:fixcolors}.  In line~3, deletion of an edge triggers the corresponding
color deletion through \(\textsc{DeleteEdgeBatch}\), so the operation remains
valid.  In line~17, if the edge has been deleted, inserting the color through
that edge is unnecessary and should be skipped.  In line~11, deleting the edge
removes the same-color conflict that would have been detected along that edge,
so skipping the operation preserves correctness.  However, this
means the conflicts considered by \(\textsc{FixColors}\) may shrink during
execution, which requires care.

Consider the procedure \(\textsc{RecolorBatch}\).  During epoch \(i+1\), after
line~3 of \cref{alg:recolorbatch}, the colors of vertices in \(S_{succ}\) are
valid with respect to the graph from epoch \(i\), excluding the edges deleted
while epoch \(i+1\) is being processed. Thus the partial coloring remains
valid. The loop in line~6 is adapted so that a
token is not propagated across a deleted edge. Intuitively, this should only improve the performance of the algorithm, and we
show that this intuition is correct.  However, some care is needed: the set of
generated tokens is not necessarily a subset of the set that would have been
generated in the absence of deletions.

This completes the \textsf{CONGEST} algorithm.  The algorithm already
satisfies the structural guarantees of \cref{thm:congest-main}, most notably
that it maintains a valid partial coloring.  It remains to prove two performance
guarantees: first, that the algorithm uses \(O(1)\) expected messages and
computation per edge in the batch; and second, that the conclusion of
\cref{lem:loground} continues to hold, so that \roundexacttheta{} rounds are
sufficient in each epoch to clear all tokens generated by the corresponding
inserted edges. Unlike the parallel setting, the
effect of edge deletions during the simulation should be accounted. To show this, we follow the analysis of \cref{sec:parallel}. 

\begin{definition}\label{def:goodleveldist}
    We call a level assignment \textbf{good} for $v$ in the distributed algorithm if it satisfies $(\Delta+1-d_\leq(v))\geq\theta(v)$ at the start of batch processing in each of the last $2\theta(v)^2$ preceding epochs in which the simulated batch insertion processes an edge incident to $v$, or all such preceding epochs if fewer exist. We evaluate each neighborhood after incorporating that epoch's acknowledged batch, as in \cref{def:goodlevelbatch}. Otherwise, we call the assignment \textbf{bad} for $v$.
\end{definition}

\begin{definition}\label{def:typicalleveldist}
    We call a level assignment \textbf{typical} for $v$ in the distributed algorithm if it satisfies the conditions of \cref{def:typicallevel} in the graph available to the recoloring computation at the start of the current processing epoch, including the batch acknowledged in that epoch. Otherwise, we call the assignment \textbf{atypical} for $v$.
\end{definition}

We define $M_v$ and $U$ in this fixed graph as before \cref{def:typicallevel}, and use its degree bounds throughout the epoch. We ignore newly inserted edges until the next epoch, while deletions can only remove edges from this graph. Thus, whenever the initial assignment is typical for $v$, the neighborhoods scanned by the initial request at $v$ and its descendants satisfy the same upper bounds throughout the epoch.

\label{par:dist-randomness}Fix a vertex $v$ and its level $l(v)$. Let $L$ denote the level assignment on $V\setminus\{v\}$, let $R$ denote the random choices made before the current processing epoch, and let $R'$ denote the fresh random choices made during that epoch, including the initial refresh choices and all subsequent recoloring attempts. We use $\mathcal W_v$ for the messages and computation charged to a potential initial recoloring request at $v$ under \cref{def:requestownership}, including all subsequent requests that inherit its ownership. The inserted edge that generated the initial request is the owner of the corresponding tokens in the distributed algorithm; distinguishing its initial requests only refines the charging of work. We write $\mathcal G_v$ and $\mathcal T_v$ for the good and typical events in \cref{def:goodleveldist,def:typicalleveldist}. Both events depend only on $L$ and the fixed update sequence.
\par

\begin{lemma}\label{lem:EVcollideCdist}
The conclusion of \cref{lem:EVcollideCbatch} remains valid for the
\textsf{CONGEST} algorithm.
\end{lemma}
\begin{proof}
We apply the proof of \cref{lem:EVcollideCbatch} to the batches processed in successive epochs. The epochs counted in \cref{def:goodleveldist} correspond to the relevant batch insertions in \cref{def:goodlevelbatch}. Within an epoch, the recoloring computation ignores new insertions, while deletions can only decrease $d_\leq(v)$. Thus, the slack bound at the start of batch processing holds throughout that epoch. Deletions introduce no new recoloring or refresh opportunities, so the collision-probability argument applies with this history window.
\par
\end{proof}

\begin{lemma}\label{lem:constantprob_dist}
    The conclusion of \cref{lem:constantprob} remains valid for the \textsf{CONGEST} algorithm.
\end{lemma}
\begin{proof}
We fix the history preceding the coloring attempt, including the palettes from which the vertices sample. Deletions during the attempt can only remove neighbors whose sampled colors are checked for conflicts. For each color sampled by a vertex, the probability of avoiding the remaining competitors is therefore at least the product bound in the proof of \cref{lem:constantprob}, evaluated with the palettes at sampling time. The same averaging argument gives success probability at least $10^{-3}$ over the fresh choices from $R'$ used in this attempt.
\par
\end{proof}

\begin{lemma}\label{lem:EVrecolor_dist}
    The conclusions of \cref{lem:EVrecolor_par} and \cref{clm:collisionExpectation} remain valid for the \textsf{CONGEST} algorithm, with goodness and typicality defined by \cref{def:goodleveldist,def:typicalleveldist}. In particular, for every typical assignment $L=\ell$ and every fixed past randomness $R=r$,
    \[
    \E_{R'}[\mathcal W_v\mid L=\ell,R=r]=O(\theta(v)),
    \]
    uniformly over the initial state and the other initial recoloring requests in the epoch. Moreover,
    \[
    \E_{L,R,R'}[\mathbf 1[\lnot\mathcal T_v]\mathcal W_v]=O(1).
    \]
\end{lemma}
\begin{proof}
We first verify \cref{clm:collisionExpectation}. Fix the history immediately before color sampling in \cref{alg:fixcolors}, and let $\mathcal A_w$ denote the palette from which each vertex $w$ samples in this attempt. As in that claim, let $IN(u)=NB_\leq(u)\cap S$, $d=|NB_>(u)|$, and $d'=|IN(u)|$, all evaluated at sampling time. Then $|\mathcal A_u|\geq d+d'+1$. Let $D\subseteq IN(u)$ be the competitors whose conflicts with $u$ are still checked after the edge deletions. The fixed update sequence determines $D$ independently of the fresh color choices.

For each $c\in\mathcal A_u$, let $g_c$ be the probability that none of the vertices in $D$ samples $c$. We keep the palettes fixed at sampling time and use the union bound to obtain
\[
\begin{aligned}
\sum_{c\in\mathcal A_u}g_c
&\geq |\mathcal A_u|-\sum_{w\in D}\frac{|\mathcal A_u\cap\mathcal A_w|}{|\mathcal A_w|}\\
&\geq |\mathcal A_u|-|D|\\
&\geq d+1.
\end{aligned}
\]
Let $o_c$ count the already-colored higher-level neighbors of $u$ whose color is $c$ and whose edges survive until propagation. Then $\sum_{c\in\mathcal A_u}o_c\leq d$. A higher-level neighbor sampling in the same attempt cannot create an additional collision request: if its sampled color matches that of $u$ along a surviving edge, \textsc{FixColors} rejects its sample and keeps its request pending. Conditional on the success of $u$, the expected number of new collision requests is therefore at most
\[
\frac{\sum_{c\in\mathcal A_u}o_cg_c}{\sum_{c\in\mathcal A_u}g_c}
\leq\frac{d}{d+1}<1.
\]

We apply these bounds after fixing $L=\ell$, $R=r$, and any history preceding the current coloring attempt, including the choices from $R'$ already used. The bounds hold over the unused choices from $R'$. By \cref{lem:constantprob_dist}, each pending request incurs at most $10^3$ attempts in expectation before success or removal. We charge these attempts using \cref{def:requestownership}. As in the proof of \cref{lem:EVrecolor_par}, averaging over the history preceding the successful attempt preserves the bound of at most one collision child in expectation. Failed attempts generate no children, so retries only increase the local cost constant and the high-level cost bound $K$ from the proof of \cref{lem:EVrecolor}.

We use the graph at the start of the epoch to bound all possible propagation paths and neighborhood scans. Deletions preserve its structural upper bounds, so we apply the recurrence $Q(i)$ from the proof of \cref{lem:EVrecolor}, with the increased cost constants just described, to every typical assignment and obtain
\[
\E_{R'}[\mathcal W_v\mid L=\ell,R=r]=O(\theta(v)).
\]
For atypical assignments, we use the same coarse bounds on the number of requests and the same constant expected number of attempts per request over the unused randomness in $R'$. The update sequence fixes the graph at the start of the epoch, so the same level-assignment tail estimates give
\[
\E_{L,R,R'}[\mathbf 1[\lnot\mathcal T_v]\mathcal W_v]=O(1).
\]
The same argument also preserves the bound conditioned on the assignment being bad, using the neighborhoods at the start of batch processing in the relevant epochs specified by \cref{def:goodleveldist}.
\par
\end{proof}

\begin{lemma}\label{lem:logroundcongest}
    The conclusion of \cref{lem:loground} remains valid for the \textsf{CONGEST} algorithm.
\end{lemma}
\begin{proof}
We use the high and low levels defined in the proof of \cref{lem:loground} and repeat its calculation of the conditional expected potential. First, condition on the bounds on the number of neighbors in each level used in that proof, evaluated in the graph available at the start of the epoch. Deletions preserve these upper bounds. Fix the history preceding a simulated coloring attempt, including the choices from $R'$ already used. By \cref{lem:constantprob_dist}, each pending vertex succeeds with probability at least $10^{-3}$. By \cref{lem:EVrecolor_dist}, a successful attempt generates at most one collision child in expectation, and every such child lies at a strictly higher level. Refreshes use the same probabilities as before and act only on surviving higher-level edges. Thus, the expected potential contributed by the children of a successful vertex satisfies the same upper bound as in \cref{lem:loground}. For the potential $\Phi_i$ defined there, we obtain
\[
\E_{R'}[\Phi_{i+1}\mid L=\ell,R,\text{history preceding step }i+1]
\leq(1-10^{-5})\Phi_i,
\]
where the expectation is over the unused choices from $R'$. The same argument clears the high levels within $O(\log n)$ simulated calls with high probability. Once those levels are clear, requests propagate only to strictly higher levels, and each remaining vertex succeeds with constant probability per attempt. Applying the same argument successively to the low levels gives the additional \roundexact{} bound from \cref{lem:loground}.
\par
\end{proof}

\begin{lemma}\label{lem:congest-base-thm}
    The conclusion of \cref{lem:parallel-base-thm} remains valid for the \textsf{CONGEST} algorithm.
\end{lemma}
\begin{proof}
We fix an inserted edge and one of its endpoints $v$, condition on $l(v)$, and use $L$, $R$, and $R'$ as defined on page~\pageref{par:dist-randomness}. Let $\mathcal P_v$ be the event that the same-color conflict on this edge generates an initial recoloring request at $v$, and let $I_v=\mathbf 1[\mathcal P_v]$. We determine this event from the colors and timestamps before processing the epoch's recolorings, so $I_v$ is fixed once we fix $L$ and $R$.

We split $\E_{L,R,R'}[I_v\mathcal W_v]$ into the same three contributions as in \cref{lem:parallel-base-thm}, using \cref{def:goodleveldist,def:typicalleveldist}. For each good and typical assignment $L=\ell$, we apply \cref{lem:EVrecolor_dist} and \cref{lem:EVcollideCdist} to obtain
\[
\begin{aligned}
\E_{R,R'}[I_v\mathcal W_v\mid L=\ell]
&=\E_R\!\left[I_v\,\E_{R'}[\mathcal W_v\mid L=\ell,R]\,\middle|\,L=\ell\right]\\
&\leq O(\theta(v))\Pr_R[\mathcal P_v\mid L=\ell]\\
&=O(1).
\end{aligned}
\]
Averaging over these assignments gives an $O(1)$ contribution. For atypical assignments, \cref{lem:EVrecolor_dist} gives
\[
\begin{aligned}
\E_{L,R,R'}[\mathbf 1[\lnot\mathcal T_v]I_v\mathcal W_v]
&\leq\E_{L,R,R'}[\mathbf 1[\lnot\mathcal T_v]\mathcal W_v]\\
&=O(1).
\end{aligned}
\]
For assignments that are typical but not good, we apply Chernoff's bound and take a union bound over the at most $2\theta(v)^2$ relevant preceding epochs in \cref{def:goodleveldist} to obtain
\[
\Pr_L[\lnot\mathcal G_v]\leq2\theta(v)^2\exp(-\Omega(\theta(v))).
\]
The uniform bound over $R'$ in \cref{lem:EVrecolor_dist} therefore gives
\[
\begin{aligned}
\E_{L,R,R'}[\mathbf 1[\lnot\mathcal G_v\cap\mathcal T_v]I_v\mathcal W_v]
&\leq O(\theta(v))\Pr_L[\lnot\mathcal G_v]\\
&=O(1).
\end{aligned}
\]
Thus, $\E_{L,R,R'}[I_v\mathcal W_v]=O(1)$ for each endpoint and each fixed $l(v)$.

For a refresh request, we condition on $L$, $R$, and the choices from $R'$ preceding its refresh choice. On a typical assignment, that choice triggers with probability $O(1/\theta(v))$, and the unused choices from $R'$ incur $O(\theta(v))$ expected messages and computation whenever it triggers. We bound the atypical contribution by the exceptional-case estimate in \cref{lem:EVrecolor_dist}. Averaging over $L,R,R'$ and $l(v)$, and adding the contributions of both endpoints and the local edge operations, gives $O(1)$ expected messages and computation per update.
\par
\end{proof}

\begin{proof}[Proof of \cref{thm:congest-main}]
The algorithm maintains the structural guarantees of
\cref{thm:congest-main}.  By \cref{lem:logroundcongest}, all tokens generated by each edge
disappear within \roundexact{} rounds with high probability. The $O(1)$ expected messages and computation per batch holds by \cref{lem:congest-base-thm}.\end{proof}

\section*{AI statement}
All the content of this paper (e.g., algorithm design and analysis) were created and written by the authors. We have used AIs in addition for further reviewing and assistance with copyediting.

\bibliographystyle{alpha}
\bibliography{library}
\appendix

\section{A Counterexample to the Intuitive Conflict Probability Bound}
\label{app:counterexample}
Here, we revisit the intuitive but incorrect bound on the probability of each insertion causing a color conflict. For context, let us use the dynamic $(1+\eps)\Delta$ coloring algorithm of Bhattacharya et al.~\cite{bhattacharya2018dynamic}. Note that this is just a simpler single-level variant, and $(\Delta+1)$ coloring is achieved by a $\Theta(\log \Delta)$-level variant of this. The simple variant suffices for exhibiting the issue on the conflict probability. In the simple $(1+\eps)\Delta$-coloring, the argument would be that since at the time of each recoloring, each node has at least $\eps\Delta$ available colors, the probability that it had choen the particular color of some  node $u$ is at most $O(1/(\eps)\Delta)$.

Let us look more closely at the argument to reflect the formal language used in the proof of Lemma 3.10 in \cite{bhattacharya2018dynamic}. The key quantity of interest is $\Pr[E_\tau]$, where $E_\tau$ is the event that for the colors assigned, we have $\chi(u) = \chi(v)$ just before the edge insertion at time $\tau$. The authors say that they bound this probability ``evaluated over the past random choices of the algorithm which the adversary fixing the order of edge insertions is oblivious to". They say let $c_u$ and $c_v$ respectively denote the colors assigned to the vertices $u$ and $v$ during at times $\tau_u$ and $\tau_v$, which are time counters that store the last “time” (edge insertion/deletion) at which $v$ are $u$ were recolored. Event $E_\tau$ occurs iff $c_u = c_v$. Without loss of generality, it is assumed that $\tau_v>\tau_u$. They condition on all the random choices made by the algorithm till just before the time $\tau_v$. Since $\tau_u < \tau_v$, this fixes the color $c_u$. At time $\tau_v$ , the vertex $v$ picks the color $c_v$ uniformly at random from a subset of colors, of size $\lambda$. They argue, and it is correct, that $\lambda=\Omega(\beta^{i-5})$. For $((1+\eps)\Delta)$, the bound is even nicer and $\lambda \geq \eps \Delta$. Hence, they conclude that $\Pr[E_\tau] \leq 1/\lambda$. In the multi-level context of $\Delta+1$ coloring translates to $O(1/\beta^{i-5})$, and in the single-level context of $(1+\eps) \Delta$-coloring, it translates to $O(1/\eps(\Delta))$.

This analysis is incorrect, as is: $\tau_v$ is a random variable, which once fixed, reveals information about the color $c_v$ being different from the colors of the other endpoints of later on insertions. Had there been an insertion later with the other endpoint having color equal to the color $c_v$ of node $v$, node $v$ would have been recolored, and $\tau_v$ would have changed. Hence, for the presumed and fixed value $\tau_v$ (for which there is no probabilistic analysis provided), the distribution of $c_v$ is not just uniform in the space of colors available at the time of sampling. Given that no recoloring has happened since $\tau_v$, the distribution gets skewed away from the colors of the other endpoints of later insertions. 

To make this point more tangible, we show a counterexample. We present a fixed (oblivious) sequence of edge insertions and deletions and show that the last update in this sequence, which is an edge insertion and fixed in advance, has probability at least $1-1/\poly(n)$ of causing a recoloring for one of its endpoints.

\paragraph{A counterexample to the per-update conflict probability bound}
Fix a constant $\varepsilon>0$, let
\[
        \Delta=n^{1/10},\qquad
        m=\lceil (1+\varepsilon)\Delta\rceil ,
\]
and let the palette be $[m]$.  The graph is initially empty.  The
vertices are initially colored independently and uniformly from $[m]$,
in the order $1,2,\ldots,n$, so that vertex $j$ is more recently colored
than vertex $i$ whenever $j>i$.

Let $C$ be a sufficiently large constant, and set
\[
        b=\lceil C m\log n\rceil,\qquad
        t=\lceil C m\log n\rceil .
\]
Since $tb=o(n)$, we may choose disjoint blocks
\[
        B_1,\ldots,B_t \subseteq \{2,\ldots,\lfloor n/2\rfloor\},
        \qquad |B_i|=b .
\]
Call a block colorful if it contains every color in $[m]$ in the initial
coloring.  By a union bound,
\[
 \Pr[\text{some }B_i\text{ is not colorful}]
 \le tm \left(1-\frac1m\right)^b
 \le tm e^{-b/m}
 \le n^{-\Omega(C)} .
\]

Let $q$ be the initial color of vertex $1$.  Consider the following fixed
update sequence. The updates are a collection of single edge insertions, each immediately followed by the deletion of that edge, and then one final edge insertion. This sequence is oblivious to the random choices of the algorithm, and
the maximum degree is at most one throughout. We will show that, regardless, the final edge has probability $1-1/\poly(n)$ of causing a color collision. 

\[
\begin{array}{ll}
\text{Phase I:}  & \text{For every }x\in B_1\cup\cdots\cup B_t,
                   \text{ insert }(1,x)\text{ and then delete }(1,x).\\[2mm]
\text{Phase II:} & \text{For }x=\lfloor n/2\rfloor+1,\ldots,n-1,
                   \text{ insert }(n,x)\text{ and then delete }(n,x).\\[2mm]
\text{Phase III:}& \text{For }i=1,\ldots,t,\text{ and every }x\in B_i,
                   \text{ insert }(n,x)\text{ and then delete }(n,x).\\[2mm]
\text{Final:}   & \text{Insert }(n,1).
\end{array}
\]

Assume first that all blocks are colorful.  During Phase I, vertex $1$
never recolors: whenever there is a conflict on $(1,x)$, vertex $x$ is
more recently colored than vertex $1$.  Hence vertex $1$ keeps color
$q$.  Moreover, after Phase I, no vertex in any $B_i$ has color $q$:
vertices that initially had color $q$ conflicted with vertex $1$ and
were recolored to a color different from $q$, while vertices with other
colors were not changed.  Since each $B_i$ was initially colorful, after
Phase I every block $B_i$ contains every color in $[m]\setminus\{q\}$,
and contains no vertex of color $q$.

During Phase II, vertex $n$ is forced to recolor with high probability.
Indeed, conditional on the initial color of $n$, the probability that no
vertex in
\[
        \{\lfloor n/2\rfloor+1,\ldots,n-1\}
\]
has that color is at most
\[
        \left(1-\frac1m\right)^{\Theta(n)}
        \ll n^{-\Omega(1)}.
\]
Thus, with probability $1-n^{-\Omega(1)}$, vertex $n$ is
recolored in Phase II, and from then on it is more recently colored than
all vertices in the blocks $B_i$.

Now consider Phase III.  At the start of a block $B_i$, if vertex $n$
has color $q$, then it will not conflict with any vertex in this block,
since $B_i$ contains no vertex of color $q$.  If vertex $n$ has some color
$c\neq q$, then $B_i$ contains a vertex of color $c$, so while scanning
the block, vertex $n$ is forced to recolor at least once.  At the first
such recoloring in the block, the graph contains only the current edge,
so the only forbidden color is the color of the other endpoint.  Since
that color is not $q$, vertex $n$ chooses color $q$ with probability
exactly
\[
        \frac1{m-1}.
\]
Therefore,
\[
 \Pr[c(n)\neq q \text{ after Phase III}]
 \le \left(1-\frac1{m-1}\right)^t
 \le \exp\!\left(-\frac{t}{m-1}\right)
 \le n^{-\Omega(C)} .
\]

Combining the above events, with probability $1-n^{-\Omega(1)}$ we have
\[
     c(1) = q = c(n)
\]
immediately before the final update.  Hence, the final ---which is the a priori fixed
insertion $(n,1)$---causes a color conflict with probability $1-n^{-\Omega(1)}$.

\end{document}